\documentclass[aps,prd,groupedaddress,10pt,nofootinbib]{revtex4-2}
\usepackage{graphicx}
\usepackage {amsmath}

\begin{document}


\title{Post-Newtonian templates for phase evolution of spherical extreme mass ratio inspirals}

\author{Norichika Sago$^1$}
\author{Ryuichi Fujita$^{2,3}$}
\author{Hiroyuki Nakano$^4$}
\affiliation{$^1$Division of General Education, Kanazawa Medical University, Kanazawa 920-0293, Japan}
\affiliation{$^2$Institute of General Education, Otemon Gakuin University, Osaka 567-8502, Japan}
\affiliation{$^3$Center for Gravitational Physics and Quantum Information, Yukawa Institute for Theoretical Physics, Kyoto University, Kyoto, 606-8502, Japan}
\affiliation{$^4$Faculty of Law, Ryukoku University, Kyoto 612-8577, Japan}

\date{\today}

\begin{abstract}
We present various post-Newtonian (PN) models for the phase evolution of compact objects
moving along quasi-spherical orbits in Kerr spacetime.
The models are derived by using the 12PN analytic formulas
of the energy, angular momentum and their averaged rates of change calculated in the framework
of the black hole perturbation theory. To examine the convergence of time-domain PN models
(TaylorT families), we evaluate the dephasing between approximants with different PN orders.
We found that the TaylorT1 model shows the best performance and the performance of the TaylorT2 model
is the next best. To evaluate the convergence of frequency-domain PN models (TaylorF families),
we evaluate the mismatch between approximants with different orders.
We found that the performance of the TaylorF2 model is comparable with the TaylorT2 model.
Although the TaylorT2 and TaylorF2 models are not so accurate as the TaylorT1 model, the fully
analytical expressions give us easy-to-handle templates and are useful to discuss effects
beyond general relativity.
\end{abstract}


\maketitle

\section{Introduction}

Extreme mass ratio inspirals (EMRIs) are important targets of space-borne gravitational
wave (GW) detectors, like Laser Interferometer Space Antenna (LISA)~\cite{LISA:2017pwj},
TianQin~\cite{TianQin:2015yph}, and Taiji~\cite{Ruan:2018tsw}. An EMRI consists of a compact
object orbiting and plunging into a massive black hole, which is expected to exist in the
center of galactic nuclei. Before plunging into the central black hole, the object emits GWs 
including a huge number of cycles. The observation and data analysis of the GWs allow us 
to measure the intrinsic parameters of the EMRI system precisely and will
provide insights into the theory of gravity, the physics and astrophysics of 
black holes~\cite{Amaro-Seoane:2010dmp, Babak:2017tow, Berry:2019wgg}.

Matched filtering is one of the promising techniques to analyze the observed GWs 
and extract the physical information from signals emitted by EMRIs accurately and
efficiently. This technique requires waveform templates representing the real signals from
EMRIs accurately. The GW templates of EMRIs are often calculated by black hole perturbation
theory because the extreme mass ratio is suitable for the small parameter
in the perturbative expansion.

In the context of black hole perturbation theory, an EMRI is modeled as a point mass
moving in Kerr spacetime, and the GWs are represented by the gravitational
perturbations induced by the point mass. The gravitational perturbations of Kerr black hole
are described by the perturbations of the Newman-Penrose variables~\cite{1962JMP.....3..566N},
which satisfy the Teukolsky equations~\cite{1972PhRvL..29.1114T,1973ApJ...185..635T} 
with the source constructed from the energy-momentum tensor for
the point mass. This means that we need to know the precise motion of the point mass
to calculate the GWs accurately by solving the Teukolsky equation.

In the test-mass limit, the motion is expressed by a geodesic in Kerr spacetime.
Taking into account the self-interaction with the own gravitational field, however,
a deviation of the mass' motion from the geodesic arises. This can be
interpreted as the ``self-force'' (see Refs.~\cite{Barack:2018yvs,2022hgwa.bookE..38P} 
and the references therein) effect.
The dominant contribution of the gravitational self-force to the mass' motion
is the time-averaged dissipative piece at the first order of the mass ratio.
It induces the losses of energy and angular momentum of the point mass and drives
the gradual inspiral into the central black hole. The dissipative effect is most
noticeable on the orbital phase and therefore the GW phase.
A phase difference of $O(1)$ significantly reduces the overlap between an observed
signal and the corresponding template, and leads to failure of the detection or
the accurate estimation of the physical parameters.
Therefore, it is important to calculate the time-averaged dissipative self-force at high
precision.

The time-averaged dissipative self-force can be rewritten in the secular changes
of three orbital parameters, the energy, azimuthal angular momentum, and Carter
parameter of the point mass. There are several works on calculating the rates of
the secular change in both numerical and analytical approaches based on the
black hole perturbation theory.
The state-of-the-art studies on the numerical approach can provide the data of
the secular evolution for general bound orbits 
in Kerr spacetime~\cite{Drasco:2005kz, Drasco:2007gn, Fujita:2009us, Fujita:2020zxe}.
Since the numerical calculation costs a lot of computational resources, however, 
the use of combination with the analytical approach is expected to reduce the cost
and improve the efficiency.
There have been several efforts in analytically calculating the secular evolution
of the orbital parameters since the 1990s
(see Refs.~\cite{Mino:1997bx, Sasaki:2003xr} and the references therein).
The post-Newtonian (PN) formulas of the secular evolution for general bound orbits
are presented in our previous works~\cite{Sago:2005fn, Ganz:2007rf, Sago:2015rpa}.

The recent advances in computers have made it possible to extend the analytical
calculation to the higher PN order. 
Several works on calculating the high PN (8th or higher PN) corrections
have been done
in Refs.~\cite{Fujita:2012cm, Fujita:2014eta, Munna:2020iju, Munna:2020juq, Munna:2023vds}.
We can also find works on the high PN formulas for the self-force correction to
the redshift invariant~\cite{Kavanagh:2015lva, Kavanagh:2016idg, Munna:2022gio, Munna:2023wce}.
The high PN calculations so far, however, are limited to equatorial orbit cases.
On the other hand, the PN calculations for non-equatorial orbits remain at the lower
order than those for equatorial cases. We find
the 2.5PN results for circular orbits with small inclination~\cite{Shibata:1994jx}
and for eccentric orbits with small inclination~\cite{Sago:2005fn}
(see also Ref.~\cite{Ganz:2007rf} with arbitrary inclination angle),
the 4PN results for generic bound orbits~\cite{Sago:2015rpa}, which have been
recently updated to 5PN order~\cite{BHPC}.
According to the major scenarios of 
EMRI formation~\cite{Amaro-Seoane:2007osp, Amaro-Seoane:2014ela},
many EMRIs are expected to be not only eccentric but also inclined.
Therefore, it is natural for one to be interested in extending the PN calculation
for non-equatorial orbits to the higher order
\footnote{In fact, at the same time as we submitted this paper,
another work on the calculation of the 12PN fluxes 
for spherical orbits~\cite{Castillo:2024isq} had appeared independently.}.

In this work, as a first step of the high PN calculation
for non-equatorial orbits, we focus on the secular evolution of a spherical orbit,
which is an orbit with a constant radius on a plane precessing around the rotation
axis of the central black hole.
Recently, a calculation of self-forced inspirals for spherical orbits
has been done in numerical approach~\cite{Lynch:2023gpu}.
In contrast, we here calculate the 12PN formulas for the secular change
of orbital parameters
through the analytic procedure based on the adiabatic approximation proposed
in Refs.~\cite{Ganz:2007rf, Sago:2015rpa}
\footnote{In Refs.~\cite{Ganz:2007rf, Sago:2015rpa}, the PN and
eccentricity-expanded formulas for generic bound orbits not only spherical
orbits are presented. Since the higher PN calculation with finite
eccentricity requires significantly more cost than the non-eccentric orbit
case, however, we restrict the calculation to the spherical orbit case
in this work.}.

To study the convergence of the PN formulas and the impact
of the higher corrections on the accuracy of the gravitational waveform,
we derive the PN templates of the phase evolution, the time-domain, TaylorT
and frequency-domain, TaylorF models~\cite{Damour:2000zb, Buonanno:2009zt}, 
for quasi-spherical orbits.
Although the PN templates in the context of the black hole
perturbation theory have been derived in Ref.~\cite{Varma:2013kna}, the work
is restricted to the circular orbit case in Schwarzschild spacetime
in exchange for the remarkably high, 22PN order.
This work presents the extension into the spherical orbit case in Kerr
spacetime, and we investigate how the convergence of the PN
formulas is influenced by the orbital inclination and the spin of the
central black hole.

This paper is organized as follows. In Sec.~\ref{sec:adiabatic_evolution},
we briefly review the spherical orbit in Kerr spacetime, and its secular
evolution driven by the dissipation due to the gravitational radiation.
In Sec.~\ref{sec:Taylor-model}, we present the TaylorT and TaylorF models
derived in a similar manner to Ref.~\cite{Varma:2013kna}. In addition, we
propose alternative choices of the TaylorT1 and TaylorT4 templates, which
appear because of increasing the number of orbital parameters.
In Sec.~\ref{sec:results}, we examine the performance of the Taylor models
for several sets of the orbital inclination and the black hole spin.
In Sec.~\ref{sec:summary}, we summarize our results.
In Appendices~\ref{sec:coeffs-EL}--\ref{sec:coeffs-dotx_T4a}, 
we present the explicit expressions of the PN
coefficients of the orbital energy and angular momentum, their time averaged
rates of change, the TaylorT and TaylorF templates, up to the 4.5PN order 
(the higher order terms are publicly available online~\cite{BHPC}).
Throughout this paper, we adopt the geometrized units with $c=G=1$ and
metric signature $(-+++)$, and the Kerr metric is described in the
Boyer–Lindquist coordinates $(t,r,\theta,\varphi)$.

\section{Adiabatic evolution of bound orbits in Kerr spacetime}
\label{sec:adiabatic_evolution}

In this section, we briefly review the generic bound geodesics in Kerr spacetime and
the formulation of the adiabatic evolution with the secular change of the orbital
parameters characterizing the geodesics.
We also present the formulation reduced to the spherical orbit case which we consider
in this paper.

\subsection{Geodesic equations in Kerr spacetime}

The equations of motion for generic orbits in Kerr spacetime with mass $M$ and Kerr parameter $a$ are
\begin{eqnarray}
\left( \frac{dr}{d\lambda} \right)^2 &=& V_r(r),
\quad
\left( \frac{d\cos\theta}{d\lambda} \right)^2
= V_\theta(\cos\theta), \cr
\frac{dt}{d\lambda} &=& V_{tr}(r) + V_{t\theta}(\cos\theta), \cr
\frac{d\varphi}{d\lambda} &=&
V_{\varphi r}(r) + V_{\varphi\theta}(\cos\theta),
\end{eqnarray}
where $\lambda$ is defined by $d/d\lambda=\Sigma d/d\tau$ ($\Sigma \equiv r^2+a^2\cos^2\theta$), and
\begin{eqnarray}
V_r(r) &\equiv& \left[ P(r) \right]^2
- \Delta \left[ r^2 + (aE-L)^2 + C\right], \cr
P(r) &\equiv& E(r^2+a^2) - aL, \quad
\Delta \equiv r^2-2Mr+a^2, \cr
V_\theta(z) &\equiv&
C - \left[ C + a^2(1-E^2) + L^2 \right] z^2 + a^2(1-E^2) z^4, \cr
V_{tr}(r) &\equiv& \frac{r^2+a^2}{\Delta}P(r),
\quad
V_{t\theta}(z) \equiv - a ( aE - L - aEz^2 ), \cr
V_{\varphi r}(r) &\equiv& \frac{aP(r)}{\Delta},
\quad
V_{\varphi\theta}(z) \equiv - aE + \frac{L}{1-z^2}.
\end{eqnarray}
Here, $z=\cos\theta$.
The geodesic motion is characterized by three parameters, $I^A=\{E, L, C\}$,
corresponding to the specific energy, angular momentum and Carter constant,
respectively. For bound orbits, we can choose an alternative parametrization
analogous to celestial mechanics, $\chi^A=\{p, e, \iota\}$, defined by
\begin{equation}
p \equiv \frac{2r_a r_p}{M(r_a+r_p)}, \quad
e \equiv \frac{r_a-r_p}{r_a+r_p}, \quad
\cos\iota \equiv \frac{L}{\sqrt{L^2+C}},
\end{equation}
where $r_p$ and $r_a$ are the values of $r$ at the periapsis and apoapsis.
We refer $p$, $e$, and $\iota$ to the non-dimensional semi-latus rectum, eccentricity, and
inclination angle, respectively.

For bound orbits, the radial and polar motions are periodic with respect
to $\lambda$. The radial and polar periods are given by
\begin{equation}
\Lambda_r = 2\int_{r_p}^{r_a} \frac{dr}{\sqrt{V_r(r)}}, \quad
\Lambda_\theta = 4\int_{0}^{z_\textrm{min}} \frac{dz}{\sqrt{V_\theta(z)}},
\end{equation}
where $z_\textrm{min}=\cos\theta_\textrm{min}$, $\theta_\textrm{min}$ is the minimal
value of $\theta$.
The radial and polar frequencies with respect to $\lambda$ are given by the
periods as
\begin{equation}
\Omega_r \equiv \frac{2\pi}{\Lambda_r}, \quad
\Omega_\theta \equiv \frac{2\pi}{\Lambda_\theta}.
\end{equation}
The temporal and azimuthal frequencies with respect to $\lambda$ can be derived as
the time average along the orbit:
\begin{eqnarray}
\Omega_t &\equiv& 
\lim_{T\to\infty} \frac{1}{2T} \int_{-T}^{T} \frac{dt}{d\lambda} d\lambda
=
\frac{1}{\Lambda_r} \int_0^{\Lambda_r} V_{tr}(r(\lambda)) d\lambda
+ \frac{1}{\Lambda_\theta} \int_0^{\Lambda_\theta}
V_{t\theta}(\cos\theta(\lambda)) d\lambda, \\
\Omega_\varphi &\equiv&
\lim_{T\to\infty} \frac{1}{2T} \int_{-T}^{T} \frac{d\varphi}{d\lambda} d\lambda
=
\frac{1}{\Lambda_r} \int_0^{\Lambda_r} V_{\varphi r}(r(\lambda)) d\lambda
+ \frac{1}{\Lambda_\theta} \int_0^{\Lambda_\theta}
V_{\varphi\theta}(\cos\theta(\lambda)) d\lambda.
\end{eqnarray}

The fundamental frequencies with respect to the time coordinate
for bound geodesics in Kerr spacetime are given by
\begin{equation}
\omega_J(E,L,C) = \frac{\Omega_J(E,L,C)}{\Omega_t(E,L,C)},
\end{equation}
where $J=\{r,\theta,\varphi\}$.

The adiabatic evolution of $\chi^A$ and orbital phase
(\textit{i.e.}, rotational angle) $\Phi$ are described by
\begin{eqnarray}
\dot{\chi}^A &=& (G^{-1})^A_B \dot{I}^B,  \label{eq:dot_chi_generic} \\
\dot{\Phi} &=& \omega_\varphi, \label{eq:dot_omega_generic}
\end{eqnarray}
where a dot ($\cdot$) over variable denotes derivative with respect to $t$
and $G^A_B\equiv (\partial I^A/\partial \chi^B)$ is the Jacobian for the
transformation from $\chi^A$ to $I^A$.

\subsection{Spherical orbits}

A spherical orbit is a bound orbit with $e=0$ (or equivalently, $r_p=r_a$), and hence
is characterized by two orbital parameters, $\{p, \iota\}$.
The parameters, $\{E,L,C\}$, are not independent for spherical orbits. In this
work, we choose $\{E, L\}$ from them as a set of independent parameters.

For the later convenience, we introduce $x\equiv(M\omega_\varphi)^{1/3}$ 
and $Y\equiv\cos\iota$ instead of $p$ and $\iota$, respectively.
Since $x$ corresponds to the characteristic velocity in the motion, it is used
as the parameter of the PN expansion.
By using the set of $\{x,Y\}$, the equations of orbital evolution,
Eqs.~\eqref{eq:dot_chi_generic} and~\eqref{eq:dot_omega_generic}, 
for spherical orbits, are reexpressed as
\begin{eqnarray}
\dot{x} &=& (G^{-1})^x_E \dot{E} + (G^{-1})^x_L \dot{L}, \label{eq:dotx} \\
\dot{Y} &=& (G^{-1})^Y_E \dot{E} + (G^{-1})^Y_L \dot{L}, \label{eq:dotY} \\
\dot{\Phi} &=& \frac{x^3}{M}, \label{eq:dotPhi}
\end{eqnarray}
where $G^A_B\equiv\partial(E,L)/\partial(x,Y)$ is the Jacobian for the transformation
from $\{x,Y\}$ to $\{E,L\}$.

The specific energy and angular momentum for a spherical orbit at the leading order of the
mass ratio are given in the form of the PN expansion
\footnote{Although $E$ and $L$ (and also $G^A_B$) are known exactly,
we treat their PN-expanded forms as in the standard PN calculation,
adjusting to the previous works~\cite{Buonanno:2009zt, Varma:2013kna}.}
as 
\begin{eqnarray}
E^{(n)}(x,Y) &=& 1 - \frac{1}{2} x^2 \sum_{k=0}^n \tilde{E}_k x^k,
\label{eq:En} \\
L^{(n)}(x,Y) &=& \frac{Y}{x} \sum_{k=0}^n \tilde{L}_k x^k.
\label{eq:Ln}
\end{eqnarray}
The superscript $(n)$ in the left-hand side of the above equations means that the formula includes the terms up
to $n$-th order terms of $x$ relative to the Newtonian order
\footnote{We consider the Newtonian order of the energy is $E^{(0)}=1-x^2/2$
although we do not employ this as the overall factor of the PN expansion.
The term like $x^n (\ln x)^k$ is treated as $O(x^n)$.}.
For example, the 2.5PN ($n=5$) energy and angular momentum are
\begin{eqnarray}
E^{(5)}(x,Y) &=& 1 - \frac{1}{2} x^2 \left\{
1 - \frac{3}{4}x^2 - \left( \frac{4}{3}-4Y \right) q x^3
- \left[ \frac{27}{8} - \left( \frac{1}{2}+Y-\frac{5}{2}Y^2 \right) q^2 \right] x^4
+ (2+6Y) q x^5
\right\}, \\
L^{(5)}(x,Y) &=&
\frac{Y}{x} \left\{ 
1 + \frac{3}{2}x^2 + \left( \frac{2}{3} - 4Y \right)x^3
+ \left[ \frac{27}{8} - \left( \frac{1}{4} + \frac{Y}{2} - \frac{7}{4}Y^2 \right) q^2 \right] x^4
- (1+6Y) x^5
\right\},
\end{eqnarray}
where $q=a/M$.
In Appendix~\ref{sec:coeffs-EL}, we show the expansion coefficients, 
$\tilde{E}_k$ and $\tilde{L}_k$, for $k\le 9$ explicitly 
(the higher order terms are publicly available online~\cite{BHPC}).

The averaged rates of change of $E$ and $L$ at the linear order of the mass ratio
are given by the black hole perturbation technique~\cite{Ganz:2007rf, Sago:2015rpa} as
\begin{eqnarray}
\dot{E}^{(n)}(x,Y) &=& -\frac{32}{5} \frac{\mu}{M^2} x^{10} 
\sum_{k=0}^n \tilde{\dot{E}}_k x^k,
\label{eq:dotEn} \\
\dot{L}^{(n)}(x,Y) &=& -\frac{32}{5} \frac{\mu}{M} x^{7} 
\sum_{k=0}^n \tilde{\dot{L}}_k x^k,
\label{eq:dotLn}
\end{eqnarray}
where $\mu$ denotes the mass of a point mass.
In the same way of Eqs.~\eqref{eq:En} and~\eqref{eq:Ln}, the superscript $(n)$ means
the truncation order. It should be noted that the coefficients, $\tilde{\dot{E}}_k$
and $\tilde{\dot{L}}_k$, include some terms with the power of $\ln x$ for
$k\ge 6$.
We have obtained the coefficients for $k\le 24$ (up to 12PN order) to date
and confirmed that they are consistent with the $e \to 0$ (circular) limit of
Ref.~\cite{Sago:2015rpa} and the $Y \to 1$ (equatorial) limit consistent with
Ref.~\cite{Fujita:2014eta}.
We have also confirmed that the 12PN formulas are consistent with the numerical
results~\cite{Fujita:2009us, Fujita:2020zxe} within numerical precision.
In Appendix~\ref{sec:coeffs-dotEL}, we show
$\tilde{\dot{E}}_k$ and $\tilde{\dot{L}}_k$ for $k\le 9$ explicitly
(the higher order terms are publicly available online~\cite{BHPC}).

\section{Post-Newtonian templates of the phase evolution}
\label{sec:Taylor-model}

The PN approximation provides various expressions of the phase evolution
with different variables and functional forms 
while keeping the same PN order~\cite{Damour:2000zb, Buonanno:2009zt, Varma:2013kna}.
In this section, we present several PN families of the phase evolution
of quasi-spherical orbits, TaylorT1, TaylorT2, TaylorT3, and TaylorT4 in the time domain
\footnote{We do not consider the TaylorEt approximant because it is shown that the
performance is not good compared to the other PN families for non-spinning, 
circular case~\cite{Buonanno:2009zt, Varma:2013kna}.
In the Kerr case, the specific energy depends on the odd power of $x$ and therefore
the expansion of $x$ with respect to $\zeta=-2E$ used in TaylorEt includes the
half-integer powers, which makes the approximant more complicated.
This is another reason for not considering the TaylorEt in this paper.},
and TaylorF1, TaylorF2 in the frequency domain, in a similar manner 
to Refs.~\cite{Damour:2000zb, Buonanno:2009zt, Varma:2013kna}.
In addition, we present subfamilies of TaylorT1 and TaylorT4, which appear because
the number of the orbital parameters increases compared to circular orbits.

\subsection{TaylorT1}

Substituting the PN formulas given by Eqs.~\eqref{eq:En}--\eqref{eq:dotLn} to
$E$, $L$, $\dot{E}$ and $\dot{L}$ in the right-hand sides of Eqs.~\eqref{eq:dotx}
and~\eqref{eq:dotY}, we obtain
\begin{eqnarray}
\dot{x} &=& (G_n^{-1})^x_E \dot{E}^{(n)} 
+ (G_n^{-1})^x_L \dot{L}^{(n)}, \label{eq:dotx_T1} \\
\dot{Y} &=& (G_n^{-1})^Y_E \dot{E}^{(n)}
+ (G_n^{-1})^Y_L \dot{L}^{(n)}, \label{eq:dotY_T1} \\
\dot{\Phi} &=& \frac{x^3}{M}, \label{eq:dotPhi_T1}
\end{eqnarray}
where $(G_n)^A_B=\partial(E^{(n)},L^{(n)})/\partial(x,Y)$.
Note that $(G_n)^A_B$ is expanded in the PN form but
the inverse, $(G_n^{-1})^A_B$, is not.
Therefore, the right-hand sides of Eqs.~\eqref{eq:dotx_T1} and~\eqref{eq:dotY_T1}
are expressed by rational functions of $x$, $\ln x$ and $Y$ although we do not
show them explicitly because of their complexity.

By analogy of the circular orbit case, we refer to the numerical solution of 
Eqs.~\eqref{eq:dotx_T1}--\eqref{eq:dotPhi_T1} as the TaylorT1 approximant
and represent the solution with respect to $\Phi$ by
$\Phi_\textrm{T1}^{(n)}$.

\subsection{TaylorT4}

The difference between the TaylorT4 and TaylorT1 approximants
\footnote{Historically, TaylorT1, TaylorT2, and TaylorT3 were developed first
in Ref.~\cite{Damour:2000zb}, 
and then TaylorT4 was introduced to compare numerical relativity simulations
for comparable-mass binary black holes in Ref.~\cite{Boyle:2007ft}.}
is whether we perform series expansion of the right-hand sides
of Eqs.~\eqref{eq:dotx_T1} and~\eqref{eq:dotY_T1} 
with respect to $x$ or not.
Expanding the right-hand sides of Eqs.~\eqref{eq:dotx_T1} and~\eqref{eq:dotY_T1}
with respect to $x$, we obtain
\begin{eqnarray}
\dot{x} &=& 
\frac{32}{5}\frac{\mu}{M^2}x^9 \sum_{k=0}^n \tilde{\dot{x}}_k x^k
\label{eq:dotx_T4} \\
&=& \frac{32}{5}\frac{\mu}{M^2}x^9 \left[
1 - \frac{743}{336}x^2 + \left( 4\pi - \frac{73Y+40}{12}q \right) x^3
+ \cdots + \left(\mbox{truncated at}\ O(x^{n+1})\right) \right], \cr
\dot{Y} &=& 
-\frac{244}{15}\frac{\mu}{M^2} x^8 q (1-Y^2) \sum_{k=0}^n \tilde{\dot{Y}}_k x^k
\label{eq:dotY_T4} \\
&=& -\frac{244}{15}\frac{\mu}{M^2} x^8 q (1-Y^2) \left[
x^3 - \frac{13}{244} q Y x^4 - \frac{10461}{1708} x^5
+ \cdots + \left(\mbox{truncated at}\ O(x^{n+1})\right) \right], \cr
\dot{\Phi} &=& \frac{x^3}{M}. \label{eq:dotPhi_T4}
\end{eqnarray}
The expansion coefficients $\tilde{\dot{x}}_k$ and $\tilde{\dot{Y}}_k$ for
$k\le 9$ are given in Appendix~\ref{sec:coeffs-dotxY_T4}.

It should be noted that, since the PN formulas of the first and second terms in the
right-hand side of Eq.~\eqref{eq:dotY_T1} are given as
\begin{eqnarray*}
(G_n^{-1})^Y_E \dot{E}^{(n)} &=&
\frac{\mu}{M^2} \left[
\frac{32}{5} Y x^8 - \frac{3502}{105} Y x^{10}
+ \left( \frac{384\pi-512q}{15} Y + \frac{152}{3} q Y^2 \right) x^{11} + \cdots
\right], \\
(G_n^{-1})^Y_L \dot{L}^{(n)} &=&
\frac{\mu}{M^2} \left[ 
- \frac{32}{5} Y x^8 + \frac{3502}{105} Y x^{10}
- \left( \frac{384\pi-512q}{15} Y + \frac{172}{5} q Y^2 + \frac{244}{15} q
\right) x^{11} + \cdots \right],
\end{eqnarray*}
the first two leading terms of them are canceling each other out. Hence, the leading
term in the PN expansion of $\dot{Y}$ is considered as the 1.5PN contribution of
the adiabatic evolution.

The TaylorT4 approximant
is obtained by solving Eqs.~\eqref{eq:dotx_T4}--\eqref{eq:dotPhi_T4} numerically.
We represent the approximant with respect to $\Phi$ by $\Phi_\textrm{T4}^{(n)}$.

\subsection{TaylorT2} \label{sec:T2}

In the circular orbit case, the TaylorT2 approximant gives $t$ and $\Phi$ as
functions of $x$ which is regarded as the parametric variable.
To obtain a similar expression for the spherical orbit case, first we derive the equation
of $Y(x)$ in the PN form as
\begin{eqnarray}
\frac{dY}{dx} &=& 
\frac{\dot{Y}}{\dot{x}} \cr
&=&
- q (1-Y^2) \left[
\frac{61}{24} x^2 - \frac{13}{96} q Y x^3 
- \left( \frac{81217}{8064} + \frac{33+15Y^2}{128} q^2 \right) x^4
+ \cdots \right].
\label{eq:dYdx_T2}
\end{eqnarray}
We can transform Eq.~\eqref{eq:dYdx_T2} into an integral form as
\begin{equation}
Y(x) = Y_I - \int_{x_I}^x dx'
q (1-Y^2) \left[
\frac{61}{24} x'^2 - \frac{13}{96} q Y x'^3
- \left( \frac{81217}{8064} + \frac{33+15Y^2}{128} q^2 \right) x'^4
+ \cdots \right], \label{eq:int_dYdx_T2}
\end{equation}
where $x_I$ is the value of $x$ at the initial time $t=0$ and $Y_I \equiv Y(x_I)$.
Considering that the integral term of Eq.~\eqref{eq:int_dYdx_T2} starts from $O(x^3)$,
we assume the PN formula of $Y(x)$ as
\begin{equation}
Y^{(n)}(x) = \tilde{Y}_0 - q (1-\tilde{Y}_0^2) \sum_{k=3}^{n} \tilde{Y}_k x^k.
\label{eq:Yn}
\end{equation}
Here the constant term, $\tilde{Y}_0$, is determined so that $Y^{(n)}(x_I)=Y_I$.
The superscript $(n)$ in the left-hand side of the above expression again 
means the truncation order as explained
below Eq.~\eqref{eq:Ln}. Also, note that the coefficients $\tilde{Y}_k$ depend on $\ln x$ in general
(later, we find that the dependence of $\ln x$ appears for $k\ge 9$).
By substituting Eq.~\eqref{eq:Yn} to $Y$ in the integral equation~\eqref{eq:int_dYdx_T2}
and comparing both sides by order of $x$, we find $\tilde{Y}_k$ iteratively.
We show the explicit formulas of $\tilde{Y}_k$ for $k\le 9$ in Appendix~\ref{sec:coeffs-Yn}.

Once we obtain the expression of $Y(x)$, the set of equations for $t(x)$ and 
$\Phi(x)$ are derived as
\begin{eqnarray}
\frac{dt}{dx} 
&=& \left( \dot{x}\bigr|_{Y=Y(x)} \right)^{-1},
\label{eq:dtdx_T2_bare} \\
\frac{d\Phi}{dx} &=&
\frac{x^3}{M} \frac{dt}{dx}.
\end{eqnarray}
Substituting Eqs.~\eqref{eq:dotx_T4} and~\eqref{eq:Yn} into the right-hand side of
Eq.~\eqref{eq:dtdx_T2_bare} and performing the PN expansion up to $O(x^n)$ relative
to the leading order, we obtain
\begin{eqnarray}
\frac{dt}{dx} 
&=&
\frac{5}{32x^9} \frac{M^2}{\mu} \left[
1 + \frac{743}{336} x^2 - \left( 4\pi - \frac{73\tilde{Y}_0+40}{12} q \right) x^3
+ \cdots + \left(\mbox{truncated at}\ O(x^{n+1})\right) \right],
\label{eq:dtdx_T2} \\
\frac{d\Phi}{dx} &=&
\frac{x^3}{M} \frac{dt}{dx}.
\label{eq:dPhidx_T2}
\end{eqnarray}
Since the right-hand sides of Eqs.~\eqref{eq:dtdx_T2} and~\eqref{eq:dPhidx_T2} are
(Laurent) polynomials of $x$ and $\ln(x)$, we can integrate them with respect to $x$
analytically and obtain the TaylorT2 approximant, which we represent by
$t_\textrm{T2}^{(n)}$ and $\Phi_\textrm{T2}^{(n)}$, as
\begin{eqnarray}
t_{\textrm{T2}}^{(n)}(x) &=&
t_{\textrm{ref}} - \frac{5}{256x^8} \frac{M^2}{\mu} 
\sum_{k=0}^n \tilde{t}_k x^k,
\label{eq:t_T2} \\
\Phi_{\textrm{T2}}^{(n)} (x) &=&
\Phi_{\textrm{ref}}
- \frac{M}{32\mu x^5} \sum_{k=0}^n \tilde{\Phi}_k^{\textrm{T2}} x^k.
\label{eq:Phi_T2}
\end{eqnarray}
The expansion coefficients, $\tilde{t}_k$ and $\tilde{\Phi}_k^{\textrm{T2}}$,
for $k\le 9$ are shown in Appendix~\ref{sec:coeffs-tPhi_T2}.
We choose constants $t_\textrm{ref}$ and $\Phi_\textrm{ref}$ so that $t(x_I)=0$
and $\Phi(x_I)=0$.

\subsection{TaylorT3}

The idea of the TaylorT3 approximant is reexpressing the TaylorT2 approximant by changing
the parametric variable from $x$ to $t$. To this end, we introduce a variable
$\Theta\equiv[\mu(t_{\textrm{ref}}-t)/(5M^2)]^{-1/8}$ and derive the inverse of Eq.~\eqref{eq:t_T2}
as a function of $\Theta$,
\begin{equation}
x(\Theta) =
\frac{\Theta}{2} \left[
1 + \frac{743}{8064} \Theta^2 + \left( \frac{73\tilde{Y}_0+40}{480} q - \frac{\pi}{10} \right) \Theta^3
+ \cdots \right].
\label{eq:x_T3}
\end{equation}
Substituting Eq.~\eqref{eq:x_T3} into Eq.~\eqref{eq:Phi_T2}, we obtain the TaylorT3 approximant
for the phase as
\begin{equation}
\Phi_{\textrm{T3}}^{(n)} (\Theta) =
\Phi_{\textrm{ref}} 
- \frac{M}{\mu\Theta^5} \sum_{k=0}^n \tilde{\Phi}_k^{\textrm{T3}} \Theta^k.
\label{eq:Phi_T3}
\end{equation}
In Appendix~\ref{sec:coeffs-Phi_T3}, we show the expression of $\tilde{\Phi}_k^{\textrm{T3}}$
for $k\le 9$.

\subsection{TaylorF1 and TaylorF2}

By using the stationary phase approximation, the waveform in the frequency domain is given
in the form of
\begin{equation}
\tilde{h}(f) = \mathcal{A} f^{-7/6} e^{i\psi(f)},
\end{equation}
where $\mathcal{A}$ corresponds to the amplitude depending on the distance and orientation
to the source~\cite{Buonanno:2009zt}. The phase is obtained by
\begin{eqnarray}
t(f) &=&
t_{\textrm{ref}} + \int_{x_f}^{x_{\textrm{ref}}} \left[
(G^{-1})^x_E \dot{E} + (G^{-1})^x_L \dot{L} \right]^{-1} dx,
\label{eq:t_f} \\
Y(f) &=& 
Y_{\textrm{ref}} + \int_{x_f}^{x_{\textrm{ref}}} \left[
(G^{-1})^Y_E \dot{E} + (G^{-1})^Y_L \dot{L} \right]
\left[ (G^{-1})^x_E \dot{E} + (G^{-1})^x_L \dot{L} \right]^{-1} dx,
\label{eq:Y_f} \\
\psi(f) &=& 2\pi f t_{\textrm{ref}} - 2 \Phi_{\textrm{ref}}
+ \frac{2}{M} \int_{x_f}^{x_{\textrm{ref}}} (x_f^3-x^3) \left[
(G^{-1})^x_E \dot{E} + (G^{-1})^x_L \dot{L} \right]^{-1} dx,
\label{eq:psi_f}
\end{eqnarray}
where $x_f^3\equiv \pi Mf$.
These equations can be reexpressed in the differential forms as
\begin{eqnarray}
\frac{dt}{df} &=& \frac{dt}{dx_f} \frac{dx_f}{df}
= - \frac{\pi M}{3x_f^2} \left[
(G^{-1})^x_E \dot{E} + (G^{-1})^x_L \dot{L} \right]^{-1},
\label{eq:dtdf} \\
\frac{dY}{df} &=& \frac{dY}{dx_f} \frac{dx_f}{df}
= - \frac{\pi M}{3x_f^2} \left[
(G^{-1})^Y_E \dot{E} + (G^{-1})^Y_L \dot{L} \right]
\left[ (G^{-1})^x_E \dot{E} + (G^{-1})^x_L \dot{L} \right]^{-1},
\label{eq:dYdf} \\
\frac{d\psi}{df} &=& 2\pi t. \label{eq:dpsidf}
\end{eqnarray}
The TaylorF1 approximant is given by solving these equations numerically
after replacing $(E, L, \dot{E}, \dot{L})$ in the right-hand sides with their
PN formulas, $(E^{(n)}, L^{(n)}, \dot{E}^{(n)}, \dot{L}^{(n)})$.
But we do not treat the TaylorF1 approximant in the rest of this paper.

By analogy from the TaylorT2 approximant in the time domain, the TaylorF2 approximant
gives $t$ and $\psi$ as functions of $f$ through $x_f$. To this end,
we first rewrite Eqs.~\eqref{eq:dtdf}--\eqref{eq:dpsidf} by expanding the right-hand sides
with respect to $x_f$ as
\begin{eqnarray}
\frac{dt}{dx_f} &=&
\frac{5}{32x_f^9} \frac{M^2}{\mu} \left[
1 + \frac{743}{336} x_f^2 - \left( 4\pi - \frac{73Y+40}{12} q \right) x_f^3
+ \cdots + \left(\mbox{truncated at}\ O(x_f^{n+1})\right) \right],
\label{eq:dtdx_F2} \\
\frac{dY}{dx_f} &=&
- q (1-Y^2) \left[
\frac{61}{24} x_f^2 - \frac{13}{96} q Y x_f^3 - \frac{80209}{8064} x_f^4
+ \cdots + \left(\mbox{truncated at}\ O(x_f^{n+1})\right) \right].
\label{eq:dYdx_F2} \\
\frac{d\psi}{dx_f} &=& 
\frac{6}{M} x_f^2 t. \label{eq:dpsidx_f}
\end{eqnarray}
Note that Eqs.~\eqref{eq:dtdx_F2} and~\eqref{eq:dYdx_F2} are equivalent to
Eqs.~\eqref{eq:dtdx_T2} and~\eqref{eq:dYdx_T2}, respectively, except for
replacing $x$ to $x_f$. Hence, we can obtain
the analytic solutions of Eqs.~\eqref{eq:dtdx_F2} and~\eqref{eq:dYdx_F2}, 
represented by $Y^{(n)}(x_f)$ and $t_\textrm{F2}^{(n)}(x_f)$,
by replacing
$x$ to $x_f$ in Eqs.~\eqref{eq:Yn} and~\eqref{eq:t_T2} as
\begin{eqnarray}
Y^{(n)}(x_f) &=& 
\tilde{Y}_0 - q(1-\tilde{Y}_0^2) \left[ \frac{61}{72} x_f^3 - \frac{13}{384} q^2 \tilde{Y}_0 x_f^4
- \frac{80209}{40320} q x_f^5 
+ \cdots \right],
\label{eq:Yn_xf} \\
t_{\textrm{F2}}^{(n)}(x_f) &=&
t_{\textrm{ref}} - \frac{5}{256x_f^8} \frac{M^2}{\mu} \left[ 
1 + \frac{743}{252} x_f^2 - \left( \frac{32}{5} \pi - \frac{146\tilde{Y}_0+80}{15} q \right) x_f^3
+ \cdots \right].
\label{eq:t_F2}
\end{eqnarray}
Substituting $t_{\textrm{F2}}^{(n)}$ into the right-hand side of Eq.~\eqref{eq:dpsidx_f}
and integrating with respect to $x_f$, we find the TaylorF2 phase as
\begin{equation}
\psi_{\textrm{F2}}^{(n)}(x_f) =
2\pi f t_c - \phi_c
+ \frac{3M}{128\mu x_f^5} \sum_{k=0}^n \tilde{\psi}_k^{\textrm{F2}} x_f^k.
\label{eq:psi_F2}
\end{equation}
The expansion coefficients, $\tilde{\psi}_k^{\textrm{F2}}$, are given in
Appendix~\ref{sec:coeffs-psi_F2}.
In this approximant, constants $t_c$ and $\phi_c$ can be chosen arbitrarily.

\subsection{Alternative TaylorT1 and TaylorT4}

As shown in Sec.~\ref{sec:T2}, we can express $Y$ as a function of $x$, $Y(x)$, 
along a quasi-spherical inspiralling orbit. By using $Y(x)$, we obtain the
parametric representation of the specific energy and the average rate of change 
with respect to $x$ as
\begin{eqnarray}
\mathcal{E}(x) &\equiv& E(x, Y(x)), \label{eq:En_parametric} \\
\mathcal{F}(x) &\equiv& \dot{E}(x, Y(x)). \label{eq:dotEn_parametric}
\end{eqnarray}
Substituting Eq.~\eqref{eq:Yn} into Eqs.~\eqref{eq:En} and~\eqref{eq:dotEn},
we can derive the PN formulas as
\begin{eqnarray}
\mathcal{E}^{(n)}(x) &=&
1 - \frac{1}{2} x^2 \sum_{k=0}^n \tilde{\mathcal{E}}_k x^k,
\label{eq:calEn} \\
\mathcal{F}^{(n)}(x) &=&
-\frac{32}{5} \frac{\mu}{M^2} x^{10} \sum_{k=0}^n \tilde{\mathcal{F}}_k x^k.
\label{eq:calFn}
\end{eqnarray}
The expansion coefficients, $\tilde{\mathcal{E}}_k$ and $\tilde{\mathcal{F}}_k$,
for $k\le 9$ are given in Appendix~\ref{sec:coeffs-calEF}.

In this representation, the adiabatic evolution of the orbital phase can be expressed in
a similar form to the circular orbit case:
\begin{eqnarray}
\dot{x} &=& \frac{\mathcal{F}^{(n)}(x)}{\mathcal{E}^{(n)\prime}(x)}, \label{eq:dotx_T1a} \\
\dot{\Phi} &=& \frac{x^3}{M}, \label{eq:dotPhi_T1a}
\end{eqnarray}
where a prime ($\prime$) denotes total derivative with respect to $x$.
We refer to the numerical solution of these equations as the alternative TaylorT1
(or TaylorT1a) approximant
and represent the solution with respect to $\Phi$ by
$\Phi_\textrm{T1a}^{(n)}$.

As in the TaylorT4 approximant, we may choose to solve Eqs.~\eqref{eq:dotx_T1a} 
and~\eqref{eq:dotPhi_T1a} after expanding the right-hand sides of Eq.~\eqref{eq:dotx_T1a}
with respect to $x$:
\begin{eqnarray}
\dot{x} &=& 
\frac{32}{5}\frac{\mu}{M^2}x^9 \sum_{k=0}^n \tilde{\dot{x}}_k^{\textrm{T4a}} x^k,
\label{eq:dotx_T4a} \\
\dot{\Phi} &=& \frac{x^3}{M}, \label{eq:dotPhi_T4a}
\end{eqnarray}
where the expansion coefficients, $\tilde{\dot{x}}_k^{\textrm{T4a}}$, are given in
Appendix~\ref{sec:coeffs-dotx_T4a}.
We name the solution in this way the alternative TaylorT4 (or TaylorT4a) approximant
and represent the solution with respect to $\Phi$ by
$\Phi_\textrm{T4a}^{(n)}$.

\section{Results} \label{sec:results}

Following Ref.~\cite{Fujita:2014eta}, we consider two kinds of EMRIs, called ``System1''
with masses $(\mu, M)=(10, 10^5)M_\odot$ and ``System2'' with masses $(\mu, M)=(10, 10^6)M_\odot$.
We assume that the observation of the GW starts when the GW frequency reaches
$f_I=4.0\times 10^{-3}$ Hz (corresponding to $x_I \sim 0.184$) for System1,
$f_I=1.8\times 10^{-3}$ Hz ($x_I \sim 0.303$) for System2, and lasts two years.
The orbital parameters and the corresponding frequency of the GW after two-year observation
are shown in Table~\ref{table:value_sys1} for System1 and Table~\ref{table:value_sys2} for
System2. The normalized radius in System1 is relatively large ($r/M \ge 18$) compared to 
System2 ($r/M \le 12$). This means that System1 is in the early stage of the inspiral,
while System2 is in the late stage.

\begin{table}[!ht]
\caption{Values of orbital parameters at start and end of observation for System1.
$x_I$, $r_I$, $Y_I$ are the values of $x$, $r$, $Y$ at start of observation, while
$x_\textrm{fin}$, $r_\textrm{fin}$, $Y_\textrm{fin}$ after two-year observation,
which are evaluated by solving the TaylorT1 equations,
Eqs.~\eqref{eq:dotx_T1}--\eqref{eq:dotPhi_T1}, at the 12PN order.
The corresponding frequency to $x_\textrm{fin}$, $f_\textrm{fin}$, is also
shown in this table.}
\label{table:value_sys1}
\begin{tabular}{cccccccc}
\hline\hline
$q$ & $x_I$ & $r_I/M$ & $Y_I$ & $x_\textrm{fin}$ & $r_\textrm{fin}/M$ 
& $Y_\textrm{fin}$ & $f_\textrm{fin}$ [Hz] \\
\hline
$0.10$ & $0.184$ & $29.7$ & $0.90$ & $0.234$ & $18.2$ & $0.900$ & $8.32\times 10^{-3}$ \\
$0.10$ & $0.184$ & $29.7$ & $0.50$ & $0.235$ & $18.1$ & $0.500$ & $8.37\times 10^{-3}$ \\
$0.10$ & $0.184$ & $29.7$ & $0.10$ & $0.235$ & $18.1$ & $0.100$ & $8.43\times 10^{-3}$ \\
$0.50$ & $0.184$ & $29.6$ & $0.90$ & $0.229$ & $19.0$ & $0.900$ & $7.75\times 10^{-3}$ \\
$0.50$ & $0.184$ & $29.7$ & $0.50$ & $0.231$ & $18.8$ & $0.498$ & $7.93\times 10^{-3}$ \\
$0.50$ & $0.184$ & $29.8$ & $0.10$ & $0.233$ & $18.6$ & $0.098$ & $8.14\times 10^{-3}$ \\
$0.90$ & $0.184$ & $29.6$ & $0.90$ & $0.225$ & $19.6$ & $0.899$ & $7.37\times 10^{-3}$ \\
$0.90$ & $0.184$ & $29.7$ & $0.50$ & $0.227$ & $19.4$ & $0.497$ & $7.58\times 10^{-3}$ \\
$0.90$ & $0.184$ & $29.8$ & $0.10$ & $0.230$ & $19.1$ & $0.096$ & $7.88\times 10^{-3}$ \\
\hline
\end{tabular}
\end{table}

\begin{table}[!ht]
\caption{Values of orbital parameters at start and end of observation for System2.
This table is corresponding to Table~\ref{table:value_sys1} for System2.}
\label{table:value_sys2}
\begin{tabular}{cccccccc}
\hline\hline
$q$ & $x_I$ & $r_I/M$ & $Y_I$ & $x_\textrm{fin}$ & $r_\textrm{fin}/M$ 
& $Y_\textrm{fin}$ & $f_\textrm{fin}$ [Hz] \\
\hline
$0.10$ & $0.303$ & $10.9$ & $0.90$ & $0.344$ & $8.45$ & $0.900$ & $2.62\times 10^{-3}$ \\
$0.10$ & $0.303$ & $10.9$ & $0.50$ & $0.346$ & $8.38$ & $0.500$ & $2.67\times 10^{-3}$ \\
$0.10$ & $0.303$ & $10.9$ & $0.10$ & $0.348$ & $8.30$ & $0.099$ & $2.72\times 10^{-3}$ \\
$0.50$ & $0.303$ & $10.8$ & $0.90$ & $0.331$ & $9.05$ & $0.900$ & $2.34\times 10^{-3}$ \\
$0.50$ & $0.303$ & $10.9$ & $0.50$ & $0.334$ & $8.98$ & $0.498$ & $2.41\times 10^{-3}$ \\
$0.50$ & $0.303$ & $11.0$ & $0.10$ & $0.340$ & $8.83$ & $0.098$ & $2.53\times 10^{-3}$ \\
$0.90$ & $0.303$ & $10.7$ & $0.90$ & $0.325$ & $9.31$ & $0.899$ & $2.22\times 10^{-3}$ \\
$0.90$ & $0.303$ & $10.9$ & $0.50$ & $0.328$ & $9.32$ & $0.497$ & $2.28\times 10^{-3}$ \\
$0.90$ & $0.303$ & $11.1$ & $0.10$ & $0.334$ & $9.23$ & $0.096$ & $2.40\times 10^{-3}$ \\
\hline
\end{tabular}
\end{table}

\subsection{Convergence of PN formulas for $\dot{E}$ and $\dot{L}$}

\begin{figure}[!ht]
\includegraphics[bb=0 0 853 518, width=0.8\linewidth]{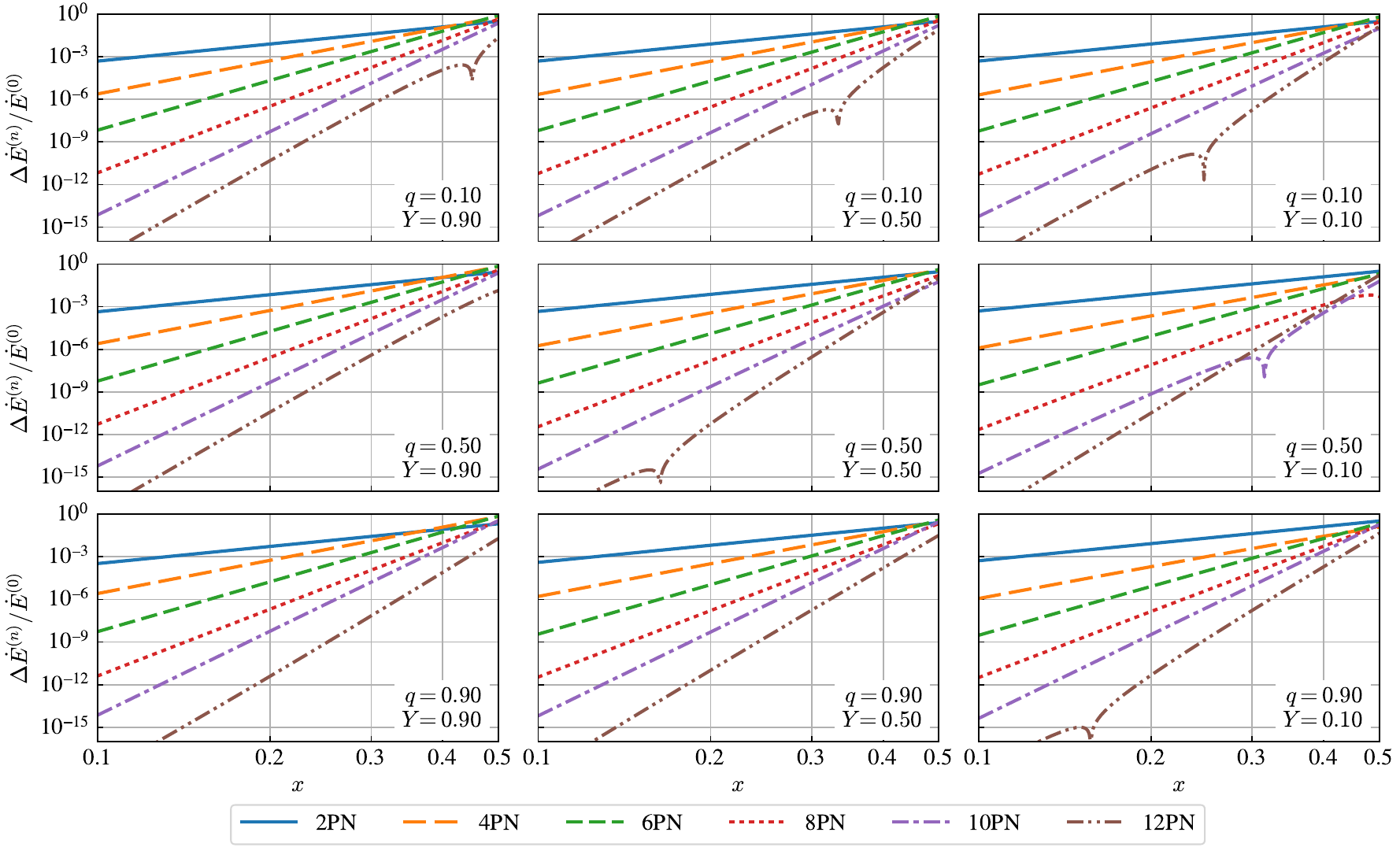}
\caption{
Contribution of each correction term in the PN expansion of $\dot{E}$.
We plot $\Delta \dot{E}^{(n)}$ normalized by the leading term, $\dot{E}^{(0)}$,
as functions of $x$ for several sets of $(q, Y)$. 
To make the plots easier to read, we show only the
2PN ($n=4$), 4PN ($n=8$), 6PN ($n=12$), 8PN ($n=16$), 10PN ($n=20$) and 12PN ($n=24$)
corrections in this figure.
Some spike bottoms are caused by the
logarithmic terms appearing in the PN formulas.
We can see that the range of convergence extends to 
larger $x$ by using higher PN order approximations.}
\label{fig:convergence-dotE}
\end{figure}

\begin{figure}[!ht]
\includegraphics[bb=0 0 853 518, width=0.8\linewidth]{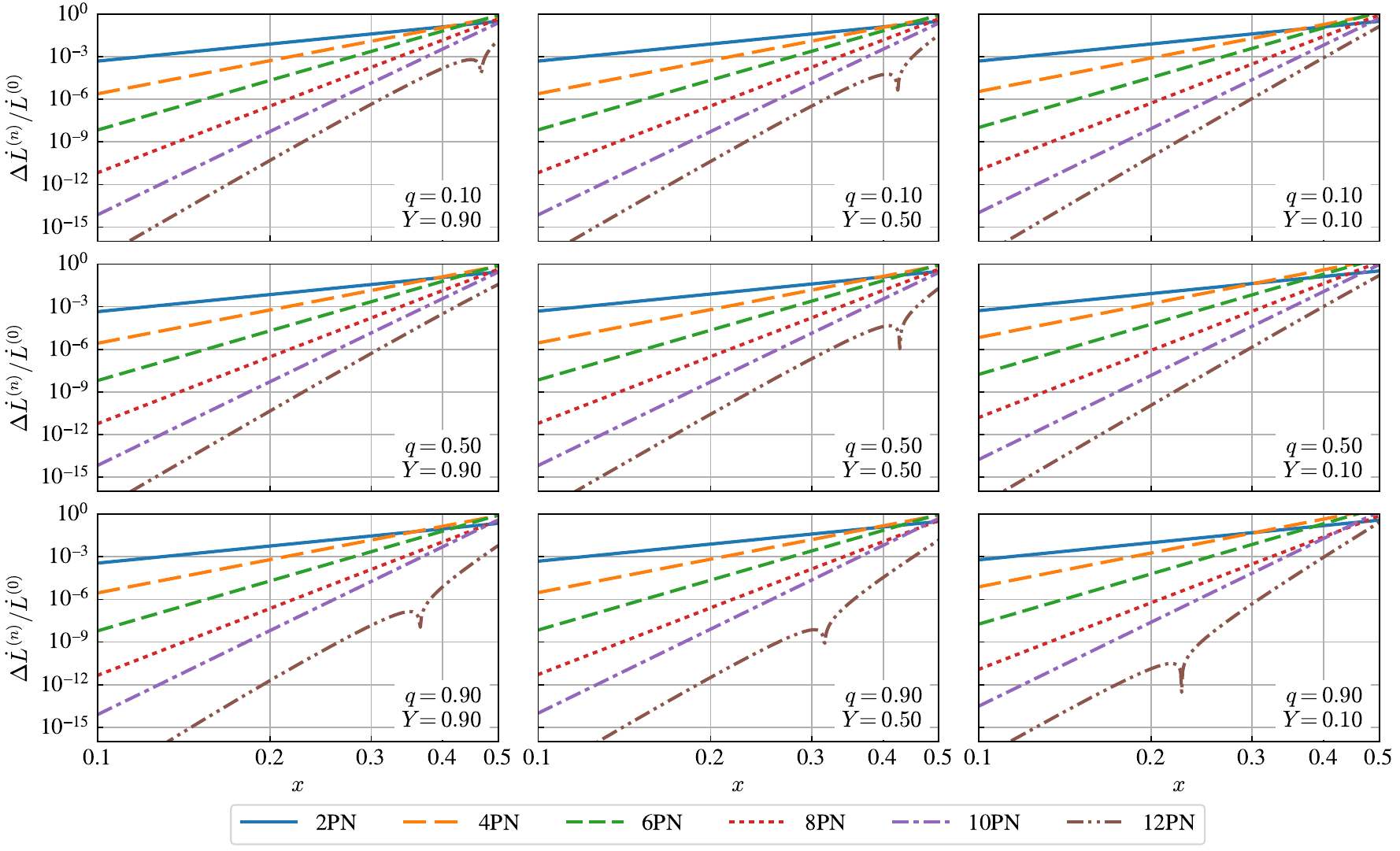}
\caption{
Contribution of each correction term in the PN expansion of $\dot{L}$.
We plot $\Delta \dot{L}^{(n)}$ normalized by the leading term, $\dot{L}^{(0)}$,
as functions of $x$ for several sets of $(q, Y)$. 
In the same reason as 
Fig.~\ref{fig:convergence-dotE}, we show only the 2PN, 4PN, 6PN, 8PN, 10PN and 12PN 
corrections in this figure. 
Some spike bottoms are caused by the
logarithmic terms appearing in the PN formulas.
We can see that the range of convergence extends to 
larger $x$ by using higher PN order approximations.}
\label{fig:convergence-dotL}
\end{figure}

To investigate the convergence of the PN formulas of $\dot{E}$ and $\dot{L}$, 
we introduce the difference between
the PN formulas with two successive orders of $x$ as
\begin{equation}
\Delta \dot{E}^{(n)} \equiv \left| \dot{E}^{(n)} - \dot{E}^{(n-1)} \right|, \quad
\Delta \dot{L}^{(n)} \equiv \left| \dot{L}^{(n)} - \dot{L}^{(n-1)} \right|.
\label{eq:PN_diff_dotEL}
\end{equation}
Since $\Delta \dot{E}^{(n)}$ ($\Delta \dot{L}^{(n)}$) 
corresponds to the $(n/2)$-th PN correction
in $\dot{E}$ ($\dot{L}$), it decreases as $n$ increases if the PN formula converges.
(See also Appendix~\ref{sec:convergence-EL} for the convergence of the PN formulas of ${E}$ and ${L}$.)

In Figs.~\ref{fig:convergence-dotE} and~\ref{fig:convergence-dotL}, 
we show $\Delta \dot{E}^{(n)}$ and $\Delta \dot{L}^{(n)}$ as functions of $x$ for
several sets of $(q, Y)$. We find some spike bottoms which are caused by the
logarithmic terms appearing in the PN formulas.
Unlike $E$ and $L$ presented in Appendix~\ref{sec:convergence-EL}, 
it is difficult to find any tendency in the convergence
of $\dot{E}$ and $\dot{L}$ depending on $q$ and $Y$. 
The convergence is not uniform comparing to $E^{(n)}$ and $L^{(n)}$, but $\Delta \dot{E}^{(n)}$ 
and $\Delta \dot{L}^{(n)}$ tend to decrease when the PN order increases. 
Therefore, we expect to obtain more accurate approximants at higher PN order.
The convergence of $\Delta \dot{E}^{(n)}$ and $\Delta \dot{L}^{(n)}$ might be improved 
by using resummation techniques~\cite{Fujita:2014eta}.
This is left to future work.

\subsection{Comparison of the time-domain Taylor approximants} \label{sec:TaylorT-results}

\begin{figure}[!ht]
\includegraphics[bb=0 0 848 440, width=0.8\linewidth]{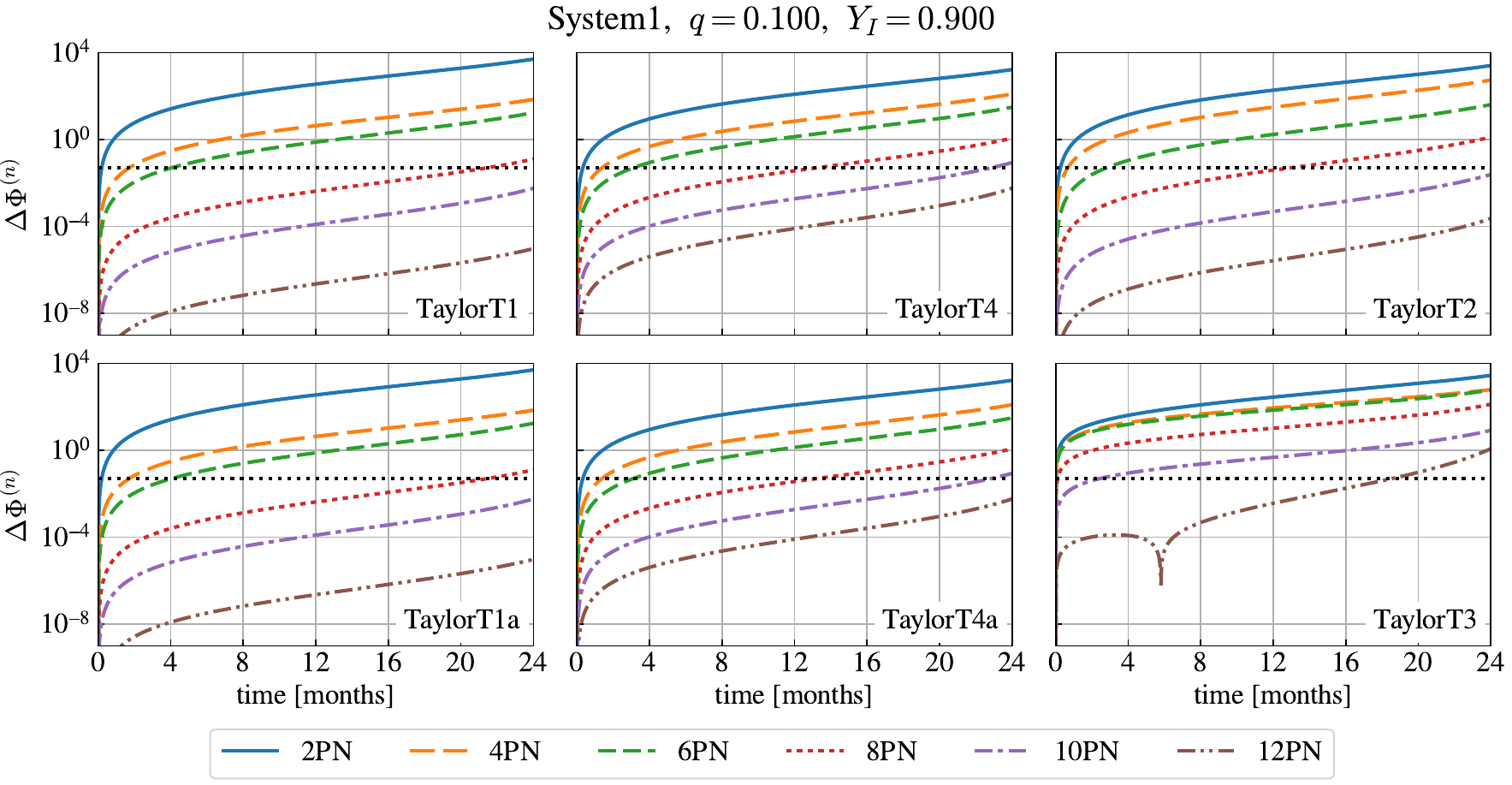}
\caption{Difference of the phase evolution, $\Delta \Phi_{X}^{(n)}$, for several TaylorT 
approximants in System1 with $q=0.1$ and $Y_I=0.9$. The horizontal dotted line denotes
$\Delta\Phi_X^{(n)}=0.05$, which corresponds to the statistical error
of the phase with the SNR of $\rho\simeq 20$.}
\label{fig:Phi_Sys1_q0.1_YI0.9}
\end{figure}

\begin{figure}[!h]
\includegraphics[bb=0 0 848 440, width=0.8\linewidth]{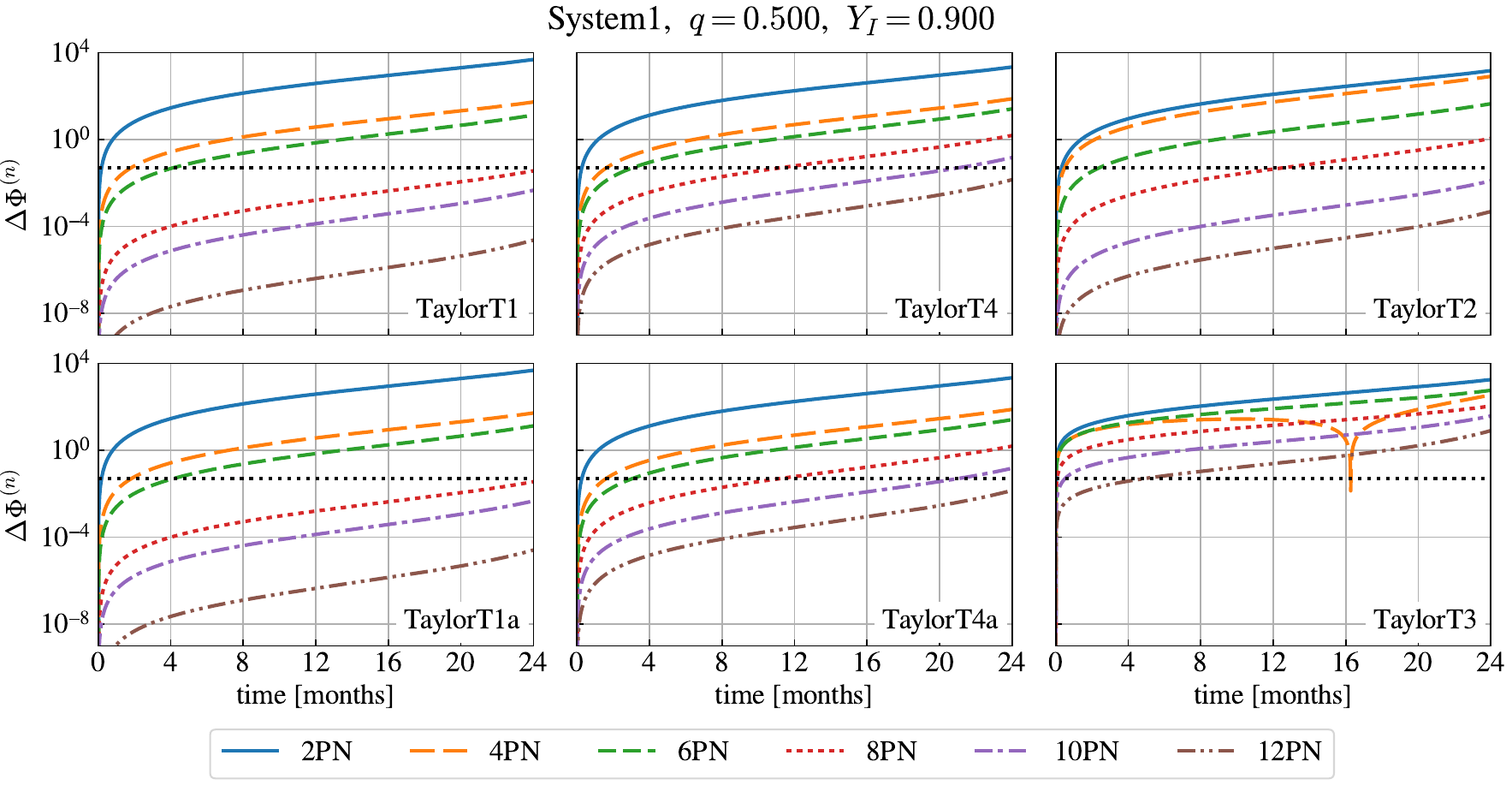}
\caption{Difference of the phase evolution, $\Delta \Phi_{X}^{(n)}$, for several TaylorT 
approximants in System1 with $q=0.5$ and $Y_I=0.9$. The horizontal dotted line denotes
$\Delta\Phi_X^{(n)}=0.05$, which corresponds to the statistical error
of the phase with the SNR of $\rho\simeq 20$.}
\label{fig:Phi_Sys1_q0.5_YI0.9}
\end{figure}

In a similar manner to Eq.~\eqref{eq:PN_diff_dotEL}, we introduce the difference in the phase between
two PN formulas with different orders of $x$ for each TaylorT approximant:
\begin{equation}
\Delta \Phi_{X}^{(n)} \equiv
\left| \Phi_{X}^{(n)} - \Phi_{X}^{(n-1)} \right|,
\end{equation}
where the index $X=$ \{T1, T2, T3, T4, T1a, T4a\} shows the name of the considered
approximant.

\begin{figure}[!ht]
\includegraphics[bb=0 0 848 440, width=0.8\linewidth]{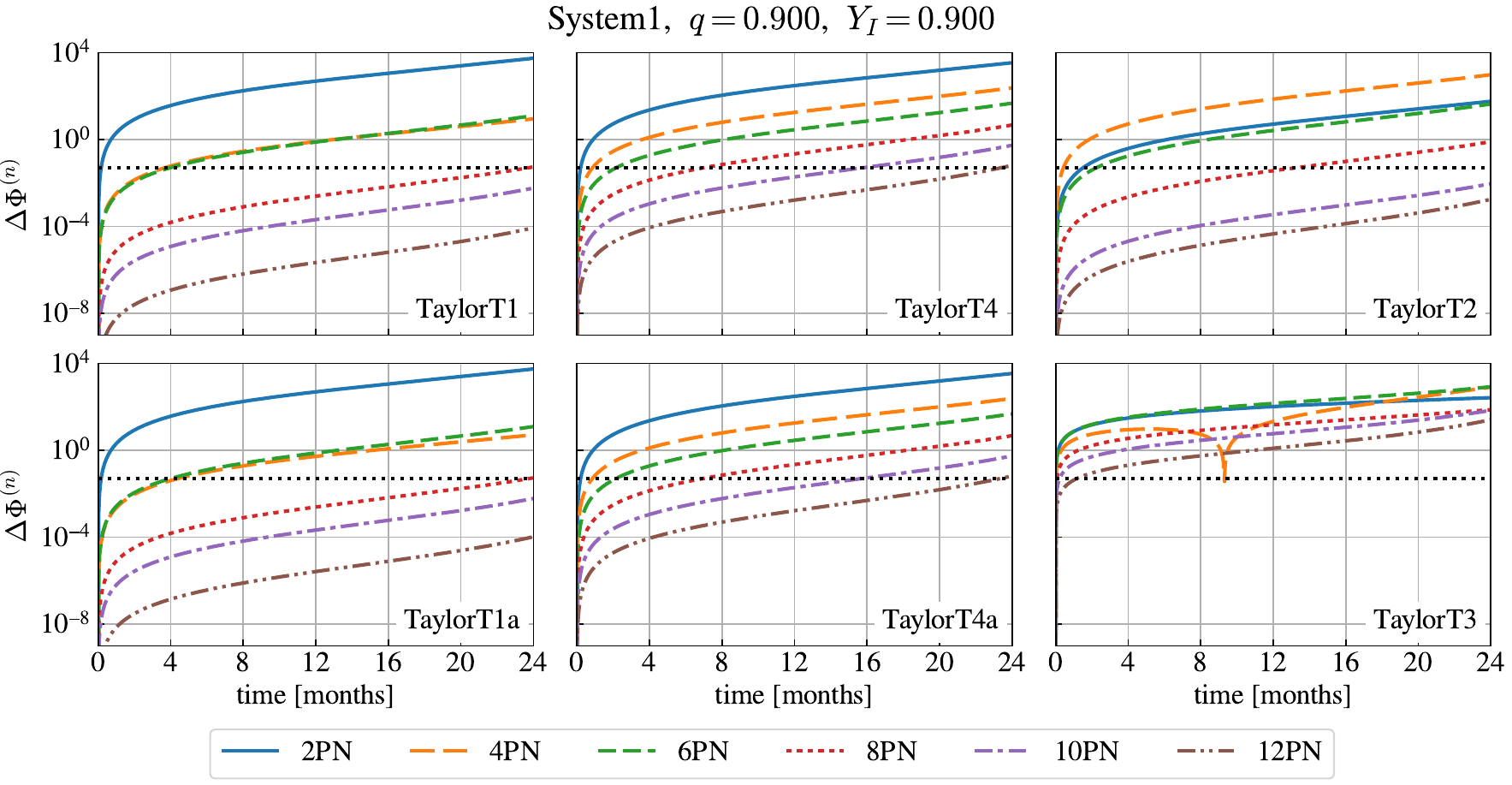}
\caption{Difference of the phase evolution, $\Delta \Phi_{X}^{(n)}$, for several TaylorT 
approximants in System1 with $q=0.9$ and $Y_I=0.9$. The horizontal dotted line denotes
$\Delta\Phi_X^{(n)}=0.05$, which corresponds to the statistical error
of the phase with the SNR of $\rho\simeq 20$.}
\label{fig:Phi_Sys1_q0.9_YI0.9}
\end{figure}
\begin{figure}[!ht]
\includegraphics[bb=0 0 848 440, width=0.8\linewidth]{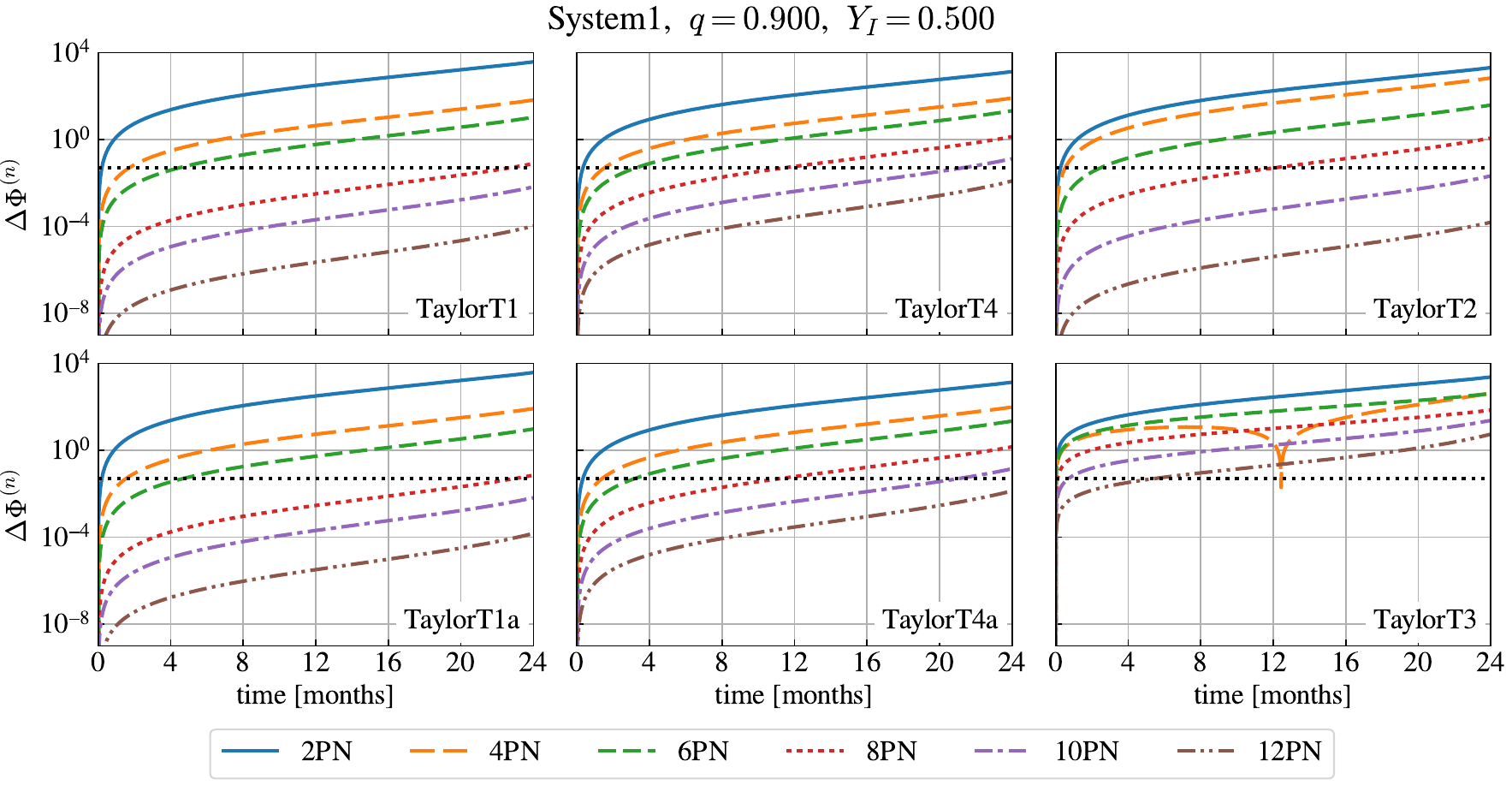}
\caption{Difference of the phase evolution, $\Delta \Phi_{X}^{(n)}$, for several TaylorT 
approximants in System1 with $q=0.9$ and $Y_I=0.5$. The horizontal dotted line denotes
$\Delta\Phi_X^{(n)}=0.05$, which corresponds to the statistical error
of the phase with the SNR of $\rho\simeq 20$.}
\label{fig:Phi_Sys1_q0.9_YI0.5}
\end{figure}
\begin{figure}[!ht]
\includegraphics[bb=0 0 848 440, width=0.8\linewidth]{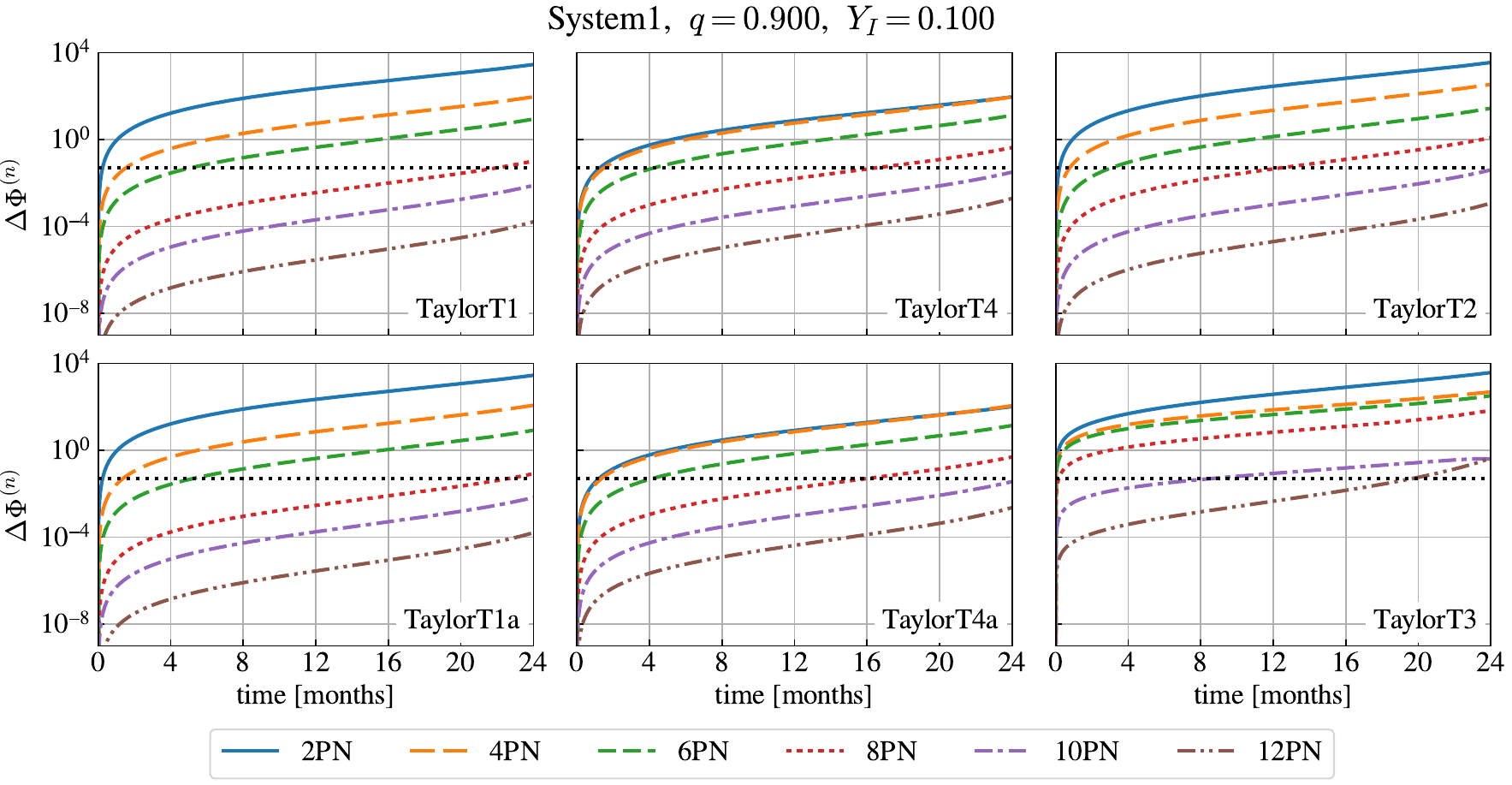}
\caption{Difference of the phase evolution, $\Delta \Phi_{X}^{(n)}$, for several TaylorT 
approximants in System1 with $q=0.9$ and $Y_I=0.1$. The horizontal dotted line denotes
$\Delta\Phi_X^{(n)}=0.05$, which corresponds to the statistical error
of the phase with the SNR of $\rho\simeq 20$.}
\label{fig:Phi_Sys1_q0.9_YI0.1}
\end{figure}

Figures~\ref{fig:Phi_Sys1_q0.1_YI0.9}--\ref{fig:Phi_Sys1_q0.9_YI0.1} show
$\Delta\Phi_X^{(n)}$ of each approximant as functions of time (in units of month)
for several sets of $(q, Y_I)$ in System1.
We find in all cases of $(q, Y_I)$ that the TaylorT1 shows best convergence.
The convergence of the TaylorT1a (TaylorT4a) is almost the same as the
TaylorT1 (TaylorT4). This suggests that the PN formula of $Y$, Eq.~\eqref{eq:Yn},
converges well.
The TaylorT2 approximant also shows good performance though the convergence is slightly
slower than the TaylorT1. The convergence of the TaylorT3 and TaylorT4 is not good
comparing to TaylorT1 and TaylorT2.
Especially, the TaylorT3 shows poor convergence comparing to the other approximants.
This trend is consistent with the result for circular orbits in Schwarzschild
spacetime~\cite{Varma:2013kna}
\footnote{For example, Ref.~\cite{Boyle:2007ft} has shown the best performance
of the TaylorT4 in the TaylorT models for comparable-mass binary black holes
(see also Ref.~\cite{Mroue:2008fu} for ineffectiveness of resummation techniques in the PN expansion).
Thus, we expected that the performance might vary depending on the situation being considered.}.
The convergence of TaylorT3 and TaylorT4 gets slightly better when $q$ and $Y_I$
are smaller. 

In all plots of Figs.~\ref{fig:Phi_Sys1_q0.1_YI0.9}--\ref{fig:Phi_Sys1_q0.9_YI0.1},
we show the line with $\Delta\Phi_X^{(n)}=0.05$ as a reference,
which corresponds to the statistical error of the phase estimated by
$\langle\Delta\Phi\rangle\sim 1/\rho$ with $\rho\simeq 20$ 
\footnote{Assuming $\rho=20$, we obtain
$\langle\Delta\Phi\rangle\sim 1/\rho=0.05$.}
($\rho$ is the signal-to-noise ratio (SNR))~\cite{Flanagan:1997kp,Lindblom:2008cm, Maggio:2021uge}.
As for the TaylorT1 approximant, the 8PN correction of the phase is less than $O(0.1)$
for two years since the GW observation starts for all cases of System1.
On the other hand, the 9th or higher PN corrections in the TaylorT2 are required to suppress
the dephasing to $O(0.1)$.
Roughly speaking, the following two higher order corrections are needed to obtain
the comparable accuracy by using the TaylorT2 approximant instead of the TaylorT1.

\begin{figure}[!ht]
\includegraphics[bb=0 0 848 440, width=0.8\linewidth]{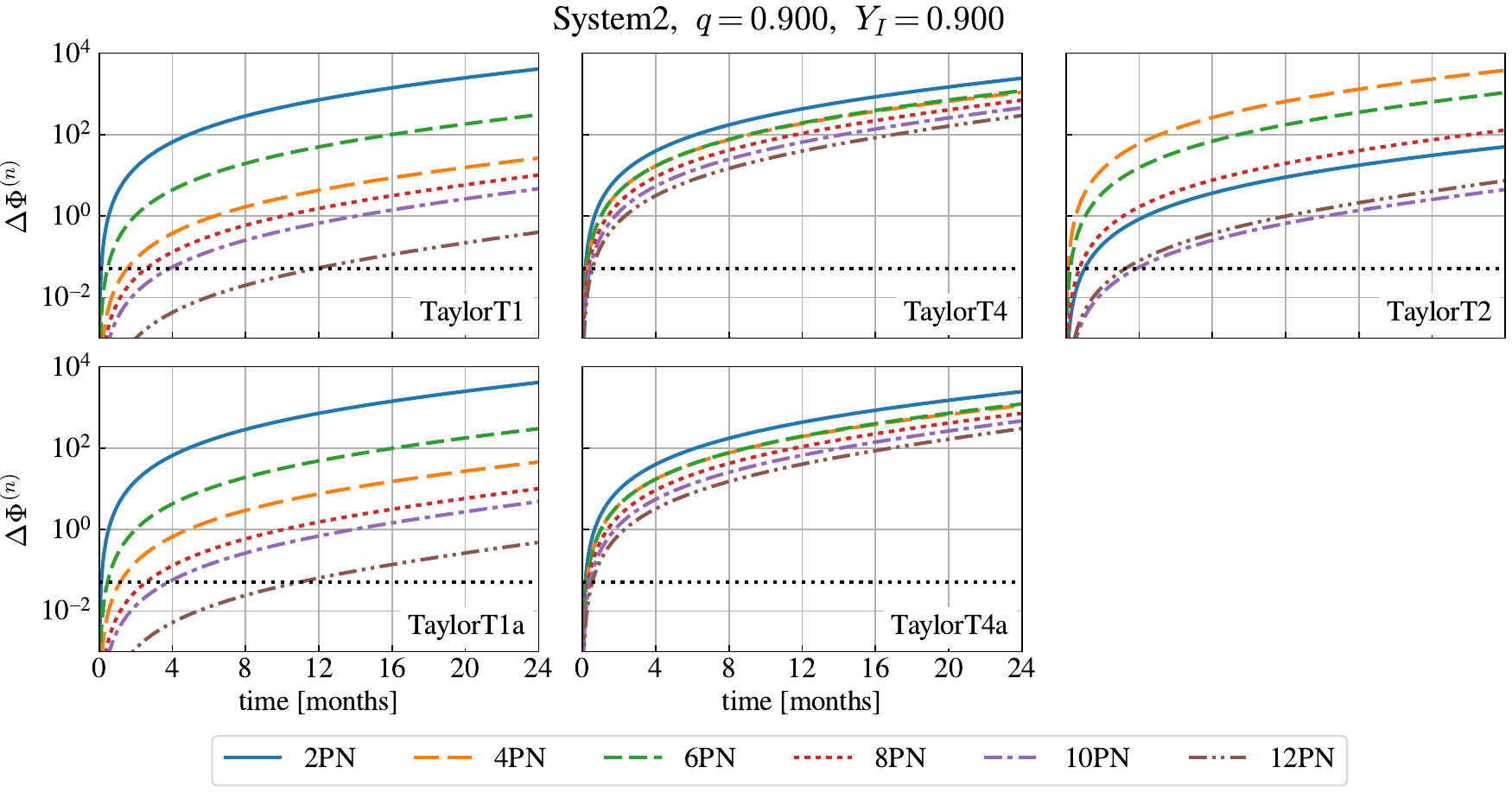}
\caption{Difference of the phase evolution, $\Delta \Phi_{X}^{(n)}$, for several TaylorT 
approximants in System2 with $q=0.9$ and $Y_I=0.9$.
The horizontal dotted line denotes $\Delta\Phi_X^{(n)}=0.05$, 
which corresponds to the statistical error of the phase with the SNR of $\rho\simeq 20$.
The plot of the TaylorT3 approximants is not shown because most of them are out of the
common range with the other approximants, $10^{-3}\le\Delta\Phi^{(n)}\le10^4$,
and the comparison does not make sense.
}
\label{fig:Phi_Sys2_q0.9_YI0.9}
\end{figure}

\begin{figure}[!ht]
\includegraphics[bb=0 0 848 440, width=0.8\linewidth]{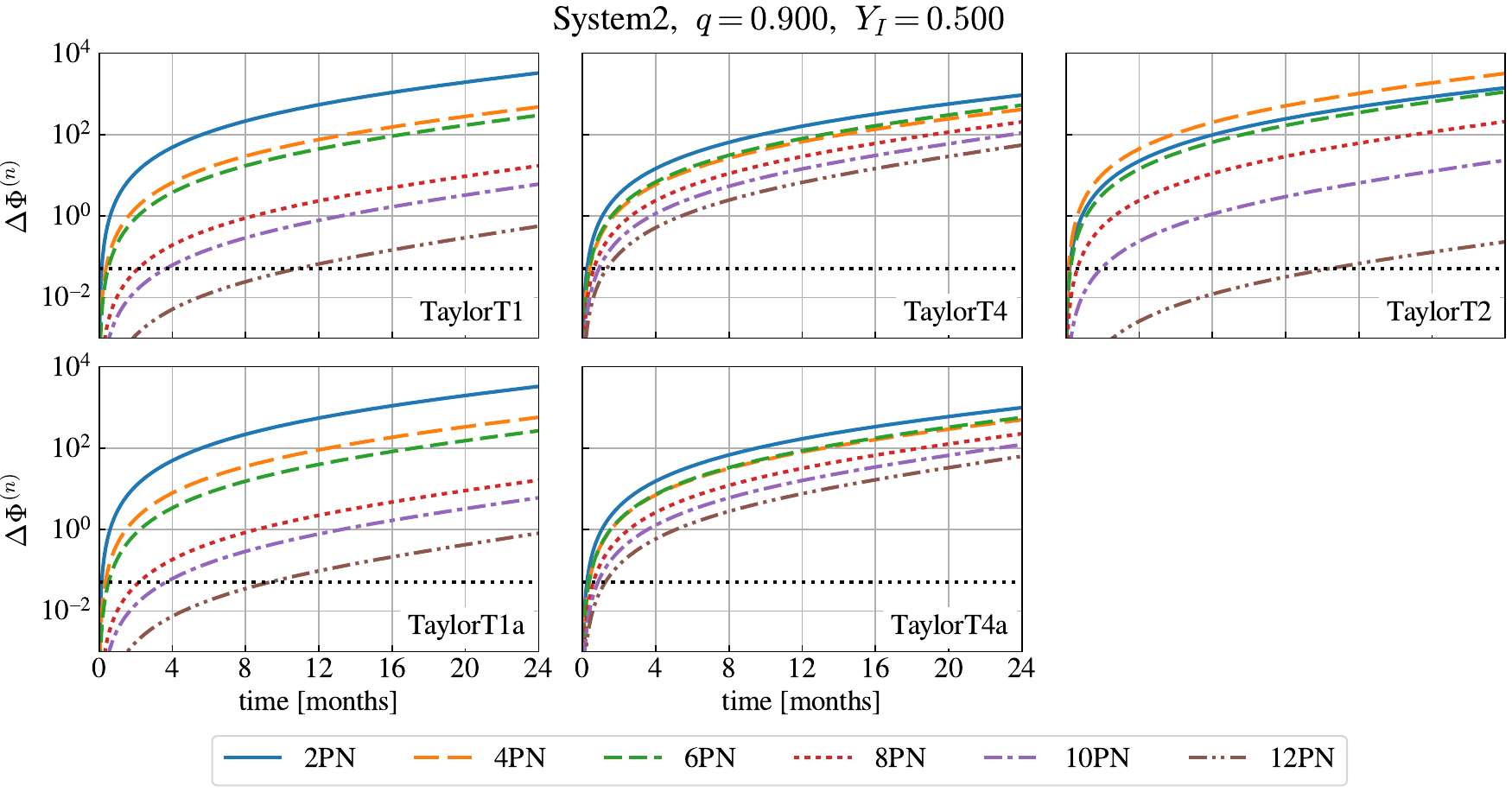}
\caption{Difference of the phase evolution, $\Delta \Phi_{X}^{(n)}$, for several TaylorT 
approximants in System2 with $q=0.9$ and $Y_I=0.5$.
The horizontal dotted line denotes $\Delta\Phi_X^{(n)}=0.05$, 
which corresponds to the statistical error of the phase with the SNR of $\rho\simeq 20$.
The plot of the TaylorT3 approximants is not shown because most of them are out of the
common range with the other approximants, $10^{-3}\le\Delta\Phi^{(n)}\le10^4$,
and the comparison does not make sense.
}
\label{fig:Phi_Sys2_q0.9_YI0.5}
\end{figure}

\begin{figure}[!ht]
\includegraphics[bb=0 0 848 440, width=0.8\linewidth]{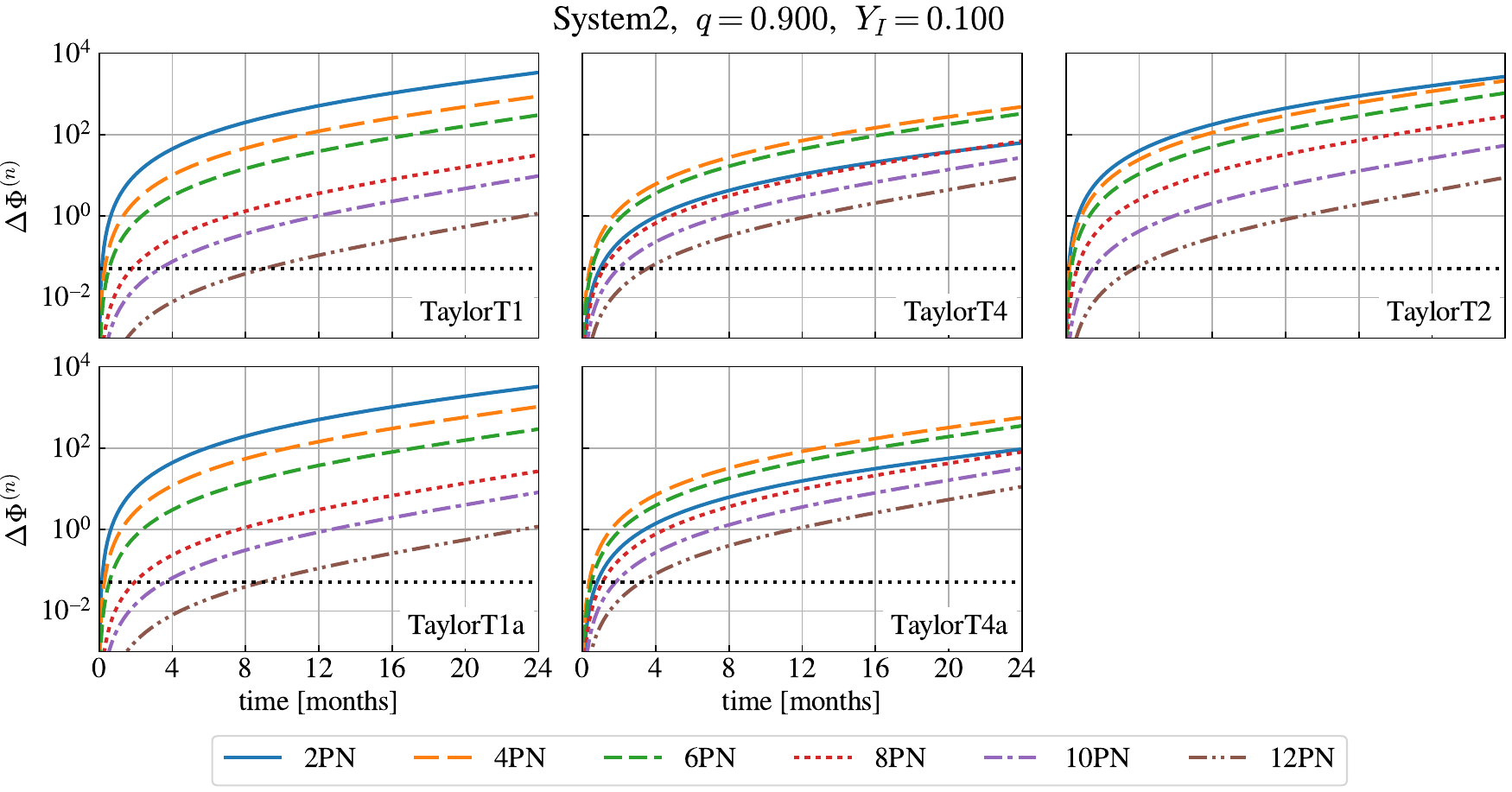}
\caption{Difference of the phase evolution, $\Delta \Phi_{X}^{(n)}$, for several TaylorT 
approximants in System2 with $q=0.9$ and $Y_I=0.1$.
The horizontal dotted line denotes $\Delta\Phi_X^{(n)}=0.05$, 
which corresponds to the statistical error of the phase with the SNR of $\rho\simeq 20$.
The plot of the TaylorT3 approximants is not shown because most of them are out of the
common range with the other approximants, $10^{-3}\le\Delta\Phi^{(n)}\le10^4$,
and the comparison does not make sense.
}
\label{fig:Phi_Sys2_q0.9_YI0.1}
\end{figure}

We show the results for System2 in
Figs.~\ref{fig:Phi_Sys2_q0.9_YI0.9}--\ref{fig:Phi_Sys2_q0.9_YI0.1}.
In the case of System2, even the 12PN correction, the highest order term we have in hand
currently, cannot keep the dephasing less than $O(0.1)$ for two years.
This is expected because System2 corresponds to the late stage of an inspiral just
before plunging into the central black hole, where the PN convergence gets worse.
The tendency of the slow convergence is more pronounced for the TaylorT3 and TaylorT4
approximants. 
Figures~\ref{fig:Phi_Sys2_q0.9_YI0.9}--\ref{fig:Phi_Sys2_q0.9_YI0.1}
do not include the results for the TaylorT3 approximant because the values are far from
the other approximants in the case of System2 and the comparison does not make sense.
The reason has been discussed for circular orbit in Schwarzschild
spacetime~\cite{Varma:2013kna}:
the slow PN convergence in the late inspiral is attributed to the pole of
$(dE/dx)^{-1}$ at the last stable orbit (LSO).
In a similar manner, the inverse of the Jacobian, $G^A_B$, appearing in
Eqs.~\eqref{eq:dotx} and~\eqref{eq:dotY}, is expected to have a pole at the LSO, 
and the PN convergence is slow near the pole.
Some techniques of resummation or factorization may be required to improve
the slow convergence around the LSO~\cite{Damour:1997ub, Damour:2008gu}.

\subsection{Performance of the frequency-domain Taylor approximants}

To discuss the convergence of the TaylorF2 approximants, it is useful to
introduce the overlap between two waveforms, $h_1(t)$ and $h_2(t)$,
\begin{equation}
\mathcal{O}(h_1|h_2) \equiv
\max_{\Delta t} \frac{\left|(h_1|h_2)\right|}
{\sqrt{(h_1|h_1)(h_2|h_2)}} ,
\end{equation}
where $\Delta t$ is the time lag between two waveforms.
The inner product, $(h_1|h_2)$, is defined by
\[
(h_1|h_2) \equiv
2 \int_{-\infty}^\infty \frac{\tilde{h}_1(f)\tilde{h}_2^*(f)}{S_n(f)}df ,
\]
with the Fourier transforms of $h_1(t)$ and $h_2(t)$, 
$\tilde{h}_1(f)$ and $\tilde{h}_2(f)$, respectively, 
and the noise power spectral density of GW detector, $S_n(f)$.
In our current calculation, we use $S_n(f)$ for LISA given in Ref.~\cite{Robson:2018ifk}.
The overlap becomes the maximum value, unity, when the two waveforms coincide perfectly
up to the overall amplitudes.

By using the overlap, we introduce the mismatch between two TaylorF2 approximants
with successive orders of $x$ as
\begin{equation}
\mathcal{M}_\textrm{F2}^{(n)} \equiv
1 - \mathcal{O}(h_\textrm{F2}^{(n)}|h_\textrm{F2}^{(n-1)}).
\end{equation}
$\mathcal{M}_\textrm{F2}^{(n)}$ corresponds to the mismatch induced by the correction
of $O(x^n)$. Hence, if the mismatch tends to decrease as the value of $n$ increase,
we diagnose that the approximants converge.

\begin{figure}
\includegraphics[bb=0 0 733 278, width=0.8\linewidth]{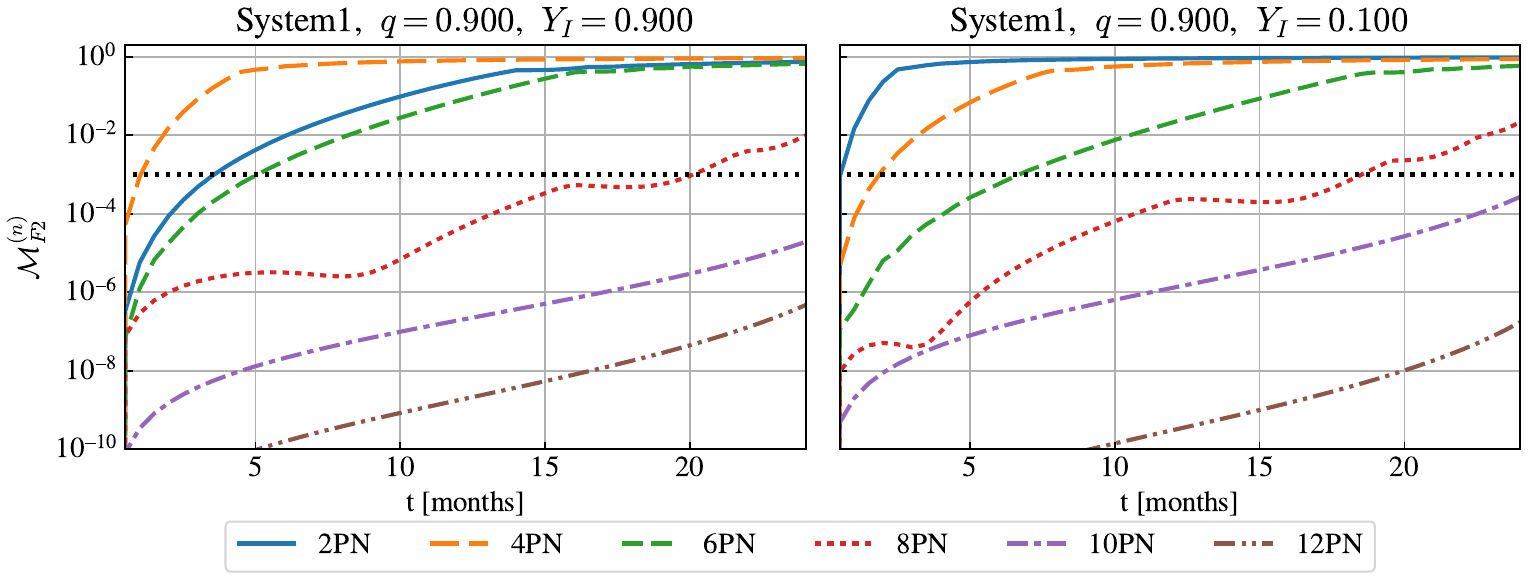}
\caption{Mismatch between different PN order TaylorF2 approximants, $\mathcal{M}_\textrm{F2}^{(n)}$
for System1 with $q=0.9$. The horizontal dotted line denotes $\mathcal{M}_\textrm{F2}^{(n)}=10^{-3}$.}
\label{fig:MM_F2_Sys1}
\end{figure}

\begin{figure}
\includegraphics[bb=0 0 733 278, width=0.8\linewidth]{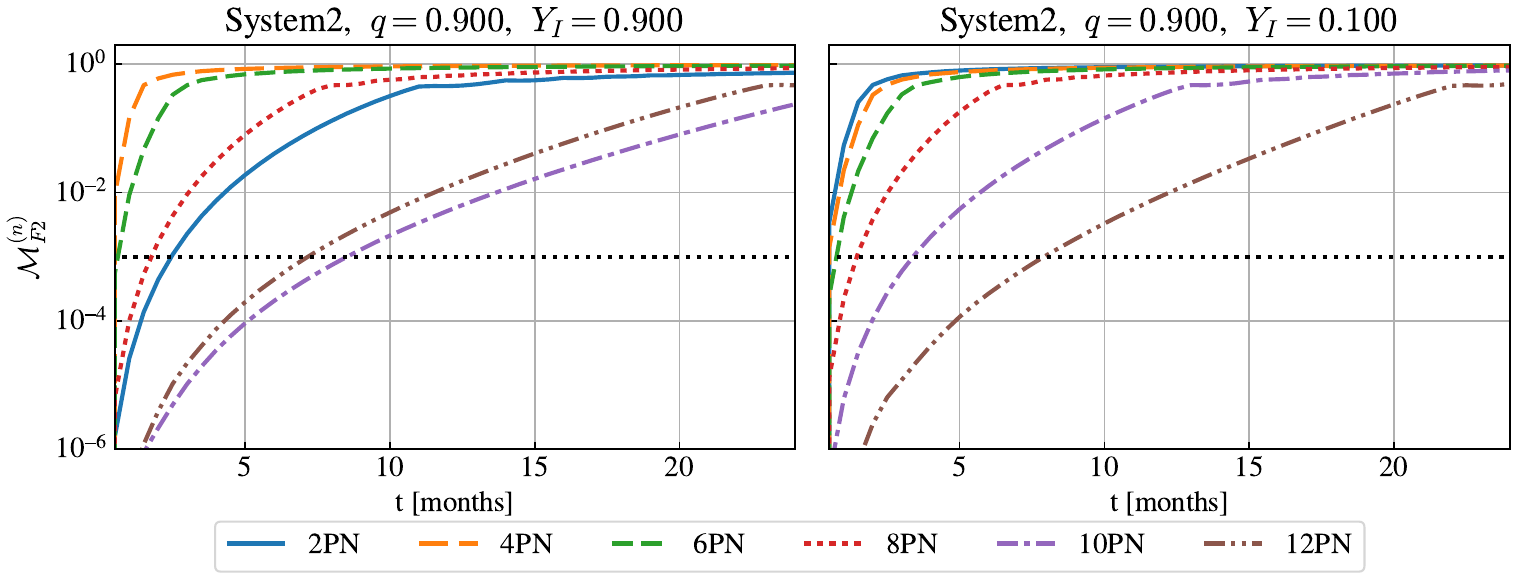}
\caption{Mismatch between different PN order TaylorF2 approximants, $\mathcal{M}_\textrm{F2}^{(n)}$
for System2 with $q=0.9$. The horizontal dotted line denotes $\mathcal{M}_\textrm{F2}^{(n)}=10^{-3}$.}
\label{fig:MM_F2_Sys2}
\end{figure}

In Figs.~\ref{fig:MM_F2_Sys1} and~\ref{fig:MM_F2_Sys2}, 
we show the mismatch as a function of observation time.
We also show the line with $\mathcal{M}_\textrm{F2}^{(n)}=10^{-3}$. This corresponds
to the limit of mismatch so that two waveforms are indistinguishable. This is estimated
by $\mathcal{M}\sim 1/(2\rho^2)$ with the SNR of $\rho\simeq 20$, 
and roughly corresponds to the dephasing
of $\Delta\Phi\sim O(0.1)$ in time-domain templates~\cite{Lindblom:2008cm, Maggio:2021uge, Datta:2021hvm}.
For System1 (early stage of inspiral), the 8PN order approximant keeps the mismatch
less than $O(0.01)$ for two-year observation, but it is not sufficient to suppress the mismatch
to $O(10^{-3})$. The 9th or higher PN order approximants are required.
The performance of the TaylorF2 approximant is similar to the TaylorT2 in the time-domain families.
As shown in Sec.~\ref{sec:TaylorT-results},
the 9th or higher PN TaylorT2 approximants are required to suppress the dephasing to
$O(0.1)$ (corresponding to a $O(10^{-3})$ mismatch in the frequency domain) for two years.
This fact may be expected from the similarity in the formulation of the TaylorT2 and
TaylorF2 approximants.

For System2 (late stage of inspiral), even the 12PN approximants, which is the highest
order we have currently, does not keep the mismatch less than $O(10^{-3})$ for two years.
This tendency is the same as we show for the phase evolution in the TaylorT families.

\section{Summary} \label{sec:summary}

In this work, we presented various PN template families for the phase evolution of
quasi-spherical EMRIs in Kerr spacetime, and examined the convergence by evaluating
the dephasing between approximants with different PN orders belonging to each family.
We found that the PN convergence slightly depends on the black hole spin, $q$, and the
initial value of the inclination, $Y_I$, and that the convergence gets slow when $q$
or $Y_I$ becomes large. This tendency implies that the rotation of the
central black hole works to slow the convergence.
In addition, the tendency suggests that we may focus on the cases with larger $q$ and
larger $Y_I$ (the case of $Y_I=0.9$ and $q=0.9$ in the current work) to discuss the
convergence of various approximants.

From the comparison of TaylorT families, we found that the TaylorT1 approximant shows
the best performance among them as shown for circular orbits in 
Schwarzschild spacetime~\cite{Varma:2013kna}. 
For early inspirals (represented by System1), the 8PN TaylorT1
template is expected to keep the dephasing less than $O(0.1)$ during two-year observation.
The alternative approximant of the TaylorT1, named TaylorT1a in this work, shows the
equivalent performance too. As for the TaylorT2 template, the corrections
at least up to the 9PN order will be required to achieve the comparable accuracy
to the TaylorT1. The convergence of the TaylorT2 approximant is slightly slower
than that of the TaylorT1/T1a. But the fully analytical expression is useful to reduce the
computational cost to construct the bank of GW templates and to discuss effects beyond
general relativity.

To examine the convergence of the TaylorF2 approximant, we evaluated the mismatch between
templates with different PN orders. For early inspirals, the corrections at least up to
9PN order will be required to keep the mismatch to $O(10^{-3})$, which corresponds to
the dephasing of $O(0.1)$ in the time domain. This result supports the expectation that
the TaylorF2 approximant has comparable performance to the TaylorT2 from the similarity
in the formulation. Since the TaylorF2 approximant is given in fully analytical
expression same as the TaylorT2, it provides an easy-to-handle method of calculating
the frequency-domain templates for EMRIs.

For late inspirals (represented by System2), the performance of all PN approximants
considered in this work gets worse than that for early inspirals. It is a natural
result because $x$, which is used as the PN parameter, becomes large as the inspiralling
object approaches the central black hole, and accordingly the convergence of the
expansion with respect to $x$ becomes slow.
Even the best template in this work, the 12PN TaylorT1 approximant, cannot keep the
dephasing to $O(0.1)$ for two years.
To satisfy $\Delta\Phi_X^{(n)} \leq 10^{-1}$ for the GW observation of System2,
higher PN (at least 14PN deducing from the Schwarzschild case~\cite{Varma:2013kna})
order calculation will be required.
There have been several efforts to improve the convergence of the energy flux by using the
factorization and resummation of the PN expansion so 
far~\cite{Damour:1997ub, Damour:2007xr, Damour:2007yf, Pan:2010hz, Isoyama:2012bx, Fujita:2014eta}.
The extension of the factorization and resummation to the spherical (and generic bound) orbits
may be studied as a future work.

In this work, we restricted binaries to the case of $e=0$. In general, it is expected
that several EMRIs in the universe have finite eccentricity when the GWs
emitted by them come into the frequency band in which LISA can detect.
The extension of the current work to generic bound orbits with three orbital parameters
is necessary to aim the GWs from EMRIs with a broad range of parameters.
The extension of the TaylorT1 and TaylorT4 families is trivial, while the derivation of
the full analytical approximants, the TaylorT2, TaylorT3, and TaylorF2, will be
more complicated than the spherical orbit case~\cite{Ganz:2007rf}.
Further investigation is needed to accomplish it.

Throughout this paper, we focus only on the dephasing in the orbital phase of
$\varphi$-direction (\textit{i.e.}, the rotation angle). 
For spherical orbits, however,
the phase with respect to $\theta$ (corresponds to the libration angle variable
of polar motion) exists, which induces GWs with other harmonics
in terms of the linear combination of orbital and polar frequencies.
For generic bound orbits in Kerr spacetime, there is another phase with respect to
the radial direction. We will work on the derivation of the PN templates for the
additional phases and harmonics in future.

\section*{Acknowledgments}

This work was supported by JSPS KAKENHI Grant Numbers
JP21H01082, JP23K20845 (N.S., R.F. and H.N.), JP21K03582 and JP23K03432 (H.N.).

\appendix

\section{Expansion coefficients of the specific energy and angular momentum}
\label{sec:coeffs-EL}

The expansion coefficients in Eqs.~\eqref{eq:En} and~\eqref{eq:Ln} up to $O(x^9)$
are given as
\begin{eqnarray*}
\tilde{E}_0 &=& 1, \quad
\tilde{E}_1 = 0, \quad
\tilde{E}_2 = -\frac{3}{4}, \quad
\tilde{E}_3 =  \left( -\frac{4}{3}+4\,Y \right) q, \cr 
\tilde{E}_4 &=& -{\frac{27}{8}}+ \left( \frac{1}{2}+Y-\frac{5}{2}\,{Y}^{2} \right) {q}^{2},
\quad
\tilde{E}_5 =  \left( 2+6\,Y \right) q, \cr 
\tilde{E}_6 &=& -{\frac{675}{64}}+ \left( {\frac{341}{36}}-{\frac {137\,Y}{6}}+{\frac {39\,{Y}^{2}}{4}} \right) {q}^{2}, \cr 
\tilde{E}_7 &=&  \left( {\frac{27}{2}}+{\frac {27\,Y}{2}} \right) q+ \left( -2-9\,Y+32\,{Y}^{2}-21\,{Y}^{3} \right) {q}^{3}, \cr 
\tilde{E}_8 &=& -{\frac{3969}{128}}+ \left( {\frac{109}{48}}-{\frac {209\,Y}{8}}+{\frac {69\,{Y}^{2}}{16}} \right) {q}^{2}+ \left( {\frac{21}{32}}+{\frac {19\,Y}{8}}-{\frac {9\,{Y}^{2}}{16}}-{\frac {101\,{Y}^{3}}{8}}+{\frac {325\,{Y}^{4}}{32}} \right) {q}^{4}, \cr 
\tilde{E}_9 &=&  \left( {\frac{225}{4}}+{\frac {135\,Y}{4}} \right) q+ \left( -{\frac{4208}{81}}+{\frac {3055\,Y}{18}}-185\,{Y}^{2}+{\frac {433\,{Y}^{3}}{6}} \right) {q}^{3}, \cr 
\tilde{L}_0 &=& 1, \quad
\tilde{L}_1 = 0, \quad 
\tilde{L}_2 = \frac{3}{2}, \quad 
\tilde{L}_3 =  \left( \frac{2}{3}-4\,Y \right) q, \cr 
\tilde{L}_4 &=& \frac{27}{8} + \left( -\frac{1}{4}-\frac{Y}{2}
+\frac{7}{4}\,{Y}^{2} \right) {q}^{2}, \quad
\tilde{L}_5 =  \left( -1-6\,Y \right) q, \cr 
\tilde{L}_6 &=& {\frac{135}{16}}+ \left( -{\frac{437}{72}}+{\frac {121\,Y}{12}}-{\frac {9\,{Y}^{2}}{8}} \right) {q}^{2}, \cr 
\tilde{L}_7 &=&  \left( -{\frac{27}{4}}-{\frac {27\,Y}{2}} \right) q+ \left( \frac{1}{2}+6\,Y-{\frac {27\,{Y}^{2}}{2}}+7\,{Y}^{3} \right) {q}^{3}, \cr 
\tilde{L}_8 &=& {\frac{2835}{128}}+ \left( -{\frac{493}{96}}+{\frac {241\,Y}{16}}+{\frac {129\,{Y}^{2}}{32}} \right) {q}^{2}+ \left( -{\frac{15}{64}}-{\frac {13\,Y}{16}}-{\frac {13\,{Y}^{2}}{32}}+{\frac {83\,{Y}^{3}}{16}}-{\frac {239\,{Y}^{4}}{64}} \right) {q}^{4}, \cr 
\tilde{L}_9 &=&  \left( -{\frac{225}{8}}-{\frac {135\,Y}{4}} \right) q+ \left( {\frac{7747}{324}}-{\frac {578\,Y}{9}}+{\frac {243\,{Y}^{2}}{4}}-{\frac {143\,{Y}^{3}}{6}} \right) {q}^{3}.
\end{eqnarray*}

\section{Expansion coefficients of $\dot{E}$ and $\dot{L}$}
\label{sec:coeffs-dotEL}

The expansion coefficients in Eqs.~\eqref{eq:dotEn} and~\eqref{eq:dotLn}
up to $O(x^9)$ are given as
\begin{eqnarray*}
\tilde{\dot{E}}_0 &=& 1, \quad
\tilde{\dot{E}}_1 = 0, \quad
\tilde{\dot{E}}_2 = -{\frac{1247}{336}}, \quad
\tilde{\dot{E}}_3 = 4 \pi +\left(-\frac{20}{3}+\frac{47 Y}{12}\right) q, \cr
\tilde{\dot{E}}_4 &=& 
-\frac{44711}{9072}+\left(-\frac{89}{96}+5 Y 
-\frac{193}{96} Y^{2}\right) q^{2}, \cr
\tilde{\dot{E}}_5 &=& 
-\frac{8191 \pi}{672}+\left(\frac{1247}{42}-\frac{11299 Y}{336}\right) q
+\left(-\frac{9}{32} Y -\frac{15}{32} Y^{3}\right) q^{3}, \cr
\tilde{\dot{E}}_6 &=& 
-\frac{3424 \ln \! \left(2\right)}{105}+\frac{16 \pi^{2}}{3}
-\frac{1712 \ln x}{105}-\frac{1712 \gamma}{105}+\frac{6643739519}{69854400}
+\left(-\frac{104}{3} \pi +\frac{143}{6} \pi  Y \right) q \cr
&&
+\left(\frac{55067}{1008}-\frac{48491}{504} Y +\frac{4793}{112} Y^{2}\right) q^{2}, \cr
\tilde{\dot{E}}_7 &=& 
-\frac{16285 \pi}{504}+\left(\frac{44711}{972}-\frac{6899 Y}{1296}\right) q 
-\frac{\left(185-1248 Y +673 Y^{2}\right) \pi  \,q^{2}}{48} \cr
&&
+\left(\frac{623}{72}-\frac{2643}{32} Y +\frac{938}{9} Y^{2}-\frac{3389}{96} Y^{3}\right) q^{3}, \cr
\tilde{\dot{E}}_8 &=&
\frac{39931 \ln \! \left(2\right)}{294}-\frac{47385 \ln \! \left(3\right)}{1568}
-\frac{1369 \pi^{2}}{126}+\frac{232597 \ln x}{4410}+\frac{232597 \gamma}{4410}
+\frac{\kappa}{2}-\frac{321516361867}{3178375200} \cr
&&
+\left[\left(\frac{1}{2}-\frac{Y^{4}}{2}\right) \Psi_B^{(0, 1)}(q)
+\left(\frac{1}{4}+\frac{3}{2} Y^{2}+\frac{1}{4} Y^{4}\right) \Psi_B^{(0, 2)}(q)
+\frac{40955 \pi}{336}-\frac{49571 \pi  Y}{336}\right] q \cr
&&
+\left[\frac{7 \kappa}{8}-\frac{2501753}{9072}+\frac{4353865 Y}{9072}
+\frac{\left(51030 \kappa -1753858\right) Y^{2}}{9072}\right] q^{2} \cr
&& +\left[\left(-\frac{3}{8}+\frac{3 Y^{4}}{8}\right) \Psi_B^{(0, 1)}(q)
+\left(\frac{3}{4}+\frac{9}{2} Y^{2}+\frac{3}{4} Y^{4}\right) \Psi_B^{(0, 2)}(q)
\right] q^{3} \cr
&&
+\left[\frac{3 \kappa}{16}+\frac{1485}{256}+\frac{67 Y}{96}+\frac{\left(1440 \kappa +10562\right) Y^{2}}{768}-\frac{3361 Y^{3}}{96}+\frac{\left(720 \kappa +11959\right) Y^{4}}{768}\right] q^{4}, \cr
\tilde{\dot{E}}_9 &=&
-\frac{6848 \pi}{105}  \left(2 \ln \! \left(2\right)+\ln x +\gamma
-\frac{265978667519}{48595599360}\right) \cr
&&
+\Biggl[\frac{18223349760-11414375280 Y}{52390800} \ln(2)
-\frac{512 \pi^{2}}{9}+\frac{54784 \ln x}{315}
+\frac{54784 \gamma}{315}-\frac{6572554559}{6548850} \cr
&&
+\frac{\left(2229519600 \pi^{2}-5698289520 \gamma -5698289520 \ln x 
+42430399563\right) Y}{52390800}\Biggr] q \cr
&& +\frac{\pi  \left(1305107-2232275 Y +977850 Y^{2}\right) q^{2}}{4032}
+\left(-\frac{240382}{567}+\frac{16357489}{12096} Y 
-\frac{326943}{224} Y^{2}+\frac{220719}{448} Y^{3}\right) q^{3} \cr
&&
+\left(-\frac{1083}{1792} Y -\frac{135}{64} Y^{2}+\frac{211}{128} Y^{3}-\frac{225}{64} Y^{4}+\frac{1095}{256} Y^{5}\right) q^{5}, \cr
\tilde{\dot{L}}_0 &=& Y, \quad
\tilde{\dot{L}}_1 = 0, \quad
\tilde{\dot{L}}_2 = -{\frac {1247\,Y}{336}}, \quad
\tilde{\dot{L}}_3 = 4\,\pi\,Y+ \left( {\frac{61}{24}}-\frac{14}{3}\,Y-\frac{5}{8}\,{Y}^{2} \right) q, \cr
\tilde{\dot{L}}_4 &=& 
-{\frac {44711\,Y}{9072}}+ \left( -{\frac {29\,Y}{16}}+\frac{7}{2}\,{Y}^{2}+\frac{3}{8}\,{Y}^{3} \right) {q}^{2}, \cr
\tilde{\dot{L}}_5 &=& 
-{\frac {8191\,\pi\,Y}{672}}+ \left( -{\frac{2661}{224}}+{\frac {1247\,Y}{56}}-{\frac {3209\,{Y}^{2}}{224}} \right) q+ \left( -{\frac{33}{128}}-{\frac {9\,{Y}^{2}}{64}}-{\frac {45\,{Y}^{4}}{128}} \right) {q}^{3}, \cr
\tilde{\dot{L}}_6 &=&
-{\frac {1712\,\ln x\,Y}{105}}+\frac{16}{3}\,{\pi}^{2}Y-{\frac {1712\,\gamma\,Y}{105}}-{\frac {3424\,\ln  \left( 2 \right) Y}{105}}+{\frac {6643739519\,Y}{69854400}} \cr
&&
+ \left( {\frac {145\,\pi}{12}}-{\frac {80\,\pi\,Y}{3}}+{\frac {15\,\pi\,{Y}^{2}}{4}} \right) q+ \left( -{\frac{305}{18}}+{\frac {181823\,Y}{4032}}-{\frac {27239\,{Y}^{2}}{672}}+{\frac {6091\,{Y}^{3}}{448}} \right) {q}^{2}, \cr
\tilde{\dot{L}}_7 &=& -{\frac {16285\,\pi\,Y}{504}}+ \left( -{\frac{80123}{6048}}+{\frac {491821\,Y}{13608}}+{\frac {322627\,{Y}^{2}}{18144}} \right) q+ \left( -{\frac {71\,\pi\,Y}{8}}+20\,\pi\,{Y}^{2}-3\,\pi\,{Y}^{3} \right) {q}^{2} \cr
&&
+ \left( -{\frac{377}{128}}+26\,Y-{\frac {10645\,{Y}^{2}}{192}}+{\frac {289\,{Y}^{3}}{8}}-{\frac {1121\,{Y}^{4}}{128}} \right) {q}^{3}, \cr
\tilde{\dot{L}}_8 &=&
{\frac {232597\,\ln x\,Y}{4410}}-{\frac {321516361867\,Y}{3178375200}}+\frac{1}{2}\,\kappa\,Y-{\frac {47385\,\ln  \left( 3 \right) Y}{1568}}-{\frac {1369\,{\pi}^{2}Y}{126}}+{\frac {232597\,\gamma\,Y}{4410}}
+{\frac {39931\,\ln  \left( 2 \right) Y}{294}} \cr
&&
+ \left[ \left( \frac{Y}{2}-\frac{1}{2}\,{Y}^{3} \right) \Psi_B^{(0,1)}(q) + \left( Y+{Y}^{3} \right) \Psi_B^{(0,2)}(q) -{\frac {16481\,\pi}{336}}+{\frac {8191\,\pi\,Y}{84}}-{\frac {3557\,\pi\,{Y}^{2}}{48}} \right] q \cr
&&
+ \left( {\frac{2661}{28}}-{\frac {8954023\,Y}{36288}}+{\frac {59\,\kappa\,Y}{16}}+{\frac {4011779\,{Y}^{2}}{18144}}-{\frac {2125175\,{Y}^{3}}{36288}}+{\frac {45\,\kappa\,{Y}^{3}}{16}} \right) {q}^{2} \cr
&&
+ \left[ \left( -\frac{3}{8}\,Y+\frac{3}{8}\,{Y}^{3} \right) \Psi_B^{(0,1)}(q) + \left( 3\,Y+3\,{Y}^{3} \right) \Psi_B^{(0,2)}(q)  \right] {q}^{3} \cr
&&
+ \left( {\frac{33}{16}}+{\frac {31\,Y}{64}}+{\frac {9\,\kappa\,Y}{8}}-{\frac {185\,{Y}^{2}}{32}}+{\frac {997\,{Y}^{3}}{64}}+{\frac {15\,\kappa\,{Y}^{3}}{8}}-{\frac {209\,{Y}^{4}}{16}}+{\frac {49\,{Y}^{5}}{32}} \right) {q}^{4}, \cr
\tilde{\dot{L}}_9 &=&  \left[ -{\frac {6848\,\pi\,Y}{105}}+ \left( -{\frac{32699}{630}}+{\frac {44512\,Y}{315}}-{\frac {5093\,{Y}^{2}}{210}} \right) q \right] \ln x-{\frac {13696\,\ln  \left( 2 \right) \pi\,Y}{105}}-{\frac {6848\,\gamma\,\pi\,Y}{105}} \cr
&&
+{\frac {265978667519\,\pi\,Y}{745113600}}+ \biggl( {\frac{22634193223}{69854400}}+{\frac {337\,{\pi}^{2}}{18}}-{\frac {65291\,\ln  \left( 2 \right) }{630}}-{\frac {32699\,\gamma}{630}}-{\frac {85229654387\,Y}{104781600}}-{\frac {416\,{\pi}^{2}Y}{9}} \cr
&&
+{\frac {89024\,\ln  \left( 2 \right) Y}{315}}+{\frac {44512\,\gamma\,Y}{315}}+{\frac {79\,{\pi}^{2}{Y}^{2}}{6}}-{\frac {5093\,\gamma\,{Y}^{2}}{210}}-{\frac {3431\,\ln  \left( 2 \right) {Y}^{2}}{70}}+{\frac {20652193823\,{Y}^{2}}{69854400}} \biggr) q \cr
&&
+ \left( -{\frac {1885\,\pi}{18}}+{\frac {1198205\,\pi\,Y}{4032}}-{\frac {29751\,\pi\,{Y}^{2}}{112}}+{\frac {38417\,\pi\,{Y}^{3}}{448}} \right) {q}^{2} \cr
&&
+ \left( {\frac{5006923}{48384}}-{\frac {9190967\,Y}{18144}}+{\frac {6411011\,{Y}^{2}}{8064}}-{\frac {55219\,{Y}^{3}}{96}}+{\frac {777995\,{Y}^{4}}{5376}} \right) {q}^{3} \cr
&&
+ \left( -{\frac{549}{896}}-{\frac {99\,Y}{64}}+{\frac {2955\,{Y}^{2}}{1792}}-{\frac {27\,{Y}^{3}}{32}}+{\frac {3\,{Y}^{4}}{32}}-{\frac {135\,{Y}^{5}}{64}}+{\frac {785\,{Y}^{6}}{256}} \right) {q}^{5},
\end{eqnarray*}
where $\gamma=0.5772156649\cdots$ is the Euler's constant, $\kappa = \sqrt{1-q^2}$, and
$\Psi_A^{(n,m)}(q)$ and $\Psi_B^{(n,m)}(q)$ are real functions defined by the polygamma
function $\psi^{(n)}(z)$ as
\begin{eqnarray}
\Psi_A^{(n,m)}(q) &\equiv&
\frac{1}{2} \left[
\psi^{(n)}\left( 3+\frac{imq}{\sqrt{1-q^2}} \right)
+ \psi^{(n)}\left( 3-\frac{imq}{\sqrt{1-q^2}} \right) \right], \\
\Psi_B^{(n,m)}(q) &\equiv&
\frac{1}{2i} \left[
\psi^{(n)}\left( 3+\frac{imq}{\sqrt{1-q^2}} \right)
- \psi^{(n)}\left( 3-\frac{imq}{\sqrt{1-q^2}} \right)
\right].
\end{eqnarray}

\section{Expansion coefficients, $\tilde{\dot{x}}_k$ and $\tilde{\dot{Y}}_k$,
in Eqs.~\eqref{eq:dotx_T4} and~\eqref{eq:dotY_T4}}
\label{sec:coeffs-dotxY_T4}

The expansion coefficients in Eqs.~\eqref{eq:dotx_T4} and~\eqref{eq:dotY_T4}
for $k\le 9$ are given as
\begin{eqnarray*}
\tilde{\dot{x}}_0 &=& 1, \quad
\tilde{\dot{x}}_1 = 0, \quad
\tilde{\dot{x}}_2 = -{\frac{743}{336}}, \quad
\tilde{\dot{x}}_3 =
4 \pi +\left(-\frac{10}{3}-\frac{73 Y}{12}\right) q, \cr
\tilde{\dot{x}}_4 &=& 
\frac{34103}{18144}+\left(-\frac{233}{96}+2 Y +\frac{527}{96} Y^{2}\right) q^{2}, \cr
\tilde{\dot{x}}_5 &=& 
-\frac{4159 \pi}{672}+\left(\frac{743}{72}-\frac{13991 Y}{336}\right) q
+\left(-\frac{9}{32} Y -\frac{15}{32} Y^{3}\right) q^{3}, \cr
\tilde{\dot{x}}_6 &=&
-\frac{3424 \ln \! \left(2\right)}{105}+\frac{16 \pi^{2}}{3}
-\frac{1712 \ln x}{105}-\frac{1712 \gamma}{105}
+\frac{16447322263}{139708800} \cr
&&
+\left(-\frac{64}{3} \pi -\frac{97}{6} \pi  Y \right) q 
+\left(\frac{41891}{4032}+\frac{17953}{1008} Y +\frac{22935}{448} Y^{2}\right) q^{2}, \cr
\tilde{\dot{x}}_7 &=&
-\frac{4415 \pi}{4032}+\left(-\frac{34103}{3024}-\frac{54595 Y}{336}\right) q
+\frac{\pi  \left(-473+672 Y +767 Y^{2}\right) q^{2}}{48} \cr
&&
+\left(\frac{95}{6}+\frac{59}{64} Y -\frac{99}{2} Y^{2}-\frac{6797}{192} Y^{3}\right) q^{3}, \cr
\tilde{\dot{x}}_8 &=&
\frac{127751 \ln \! \left(2\right)}{1470}-\frac{47385 \ln \! \left(3\right)}{1568}
-\frac{361 \pi^{2}}{126}+\frac{124741 \ln x}{4410}
+\frac{124741 \gamma}{4410}+\frac{\kappa}{2}+\frac{3971984677513}{25427001600} \cr
&&
+\left[\left(\frac{1}{2}-\frac{Y^{4}}{2}\right) \Psi_B^{(0,1)}(q)
+\left(\frac{1}{4}+\frac{3}{2} Y^{2}+\frac{1}{4} Y^{4}\right) \Psi_B^{(0,2)}(q)
+\frac{20795 \pi}{504}-\frac{5429 \pi  Y}{28}\right] q \cr
&&
+\left[\frac{7 \kappa}{8}-\frac{26297639}{145152}+\frac{207683 Y}{648}
+\frac{\left(816480 \kappa +85019873\right) Y^{2}}{145152}\right] q^{2} \cr
&&
+\left[\left(-\frac{3}{8}+\frac{3 Y^{4}}{8}\right) \Psi_B^{(0,1)}(q)
+\left(\frac{3}{4}+\frac{9}{2} Y^{2}+\frac{3}{4} Y^{4}\right) \Psi_B^{(0,2)}(q)
\right] q^{3} \cr
&&
+\left[\frac{3 \kappa}{16}+\frac{1445}{256}-\frac{1517 Y}{192}
+\frac{\left(1440 \kappa -9554\right) Y^{2}}{768}+\frac{4829 Y^{3}}{192}
+\frac{\left(720 \kappa +8099\right) Y^{4}}{768}\right] q^{4}, \cr
\tilde{\dot{x}}_9 &=&
-\frac{6848 \pi}{105}  \left(\ln x+2 \ln \! \left(2\right)+\gamma 
-\frac{343801320119}{48595599360}\right) \cr
&&
+\Biggl[\frac{\left(100228423680+45360120960 Y \right) \ln \! \left(2\right)}{419126400}
+\frac{\left(50114211840+22751245440 Y \right) \ln x}{419126400} \cr
&&
-\frac{352 \pi^{2}}{9}+\frac{37664 \gamma}{315}-\frac{16240238743}{19051200}
+\frac{\left(-4517251200 \pi^{2}+22751245440 \gamma
-578438815461\right) Y}{419126400}\Biggr] q \cr
&&
+\frac{\left(482588+43163 Y +812745 Y^{2}\right) \pi  \,q^{2}}{4032}
+\left(-\frac{1672145}{36288}+\frac{18305605}{48384} Y 
-\frac{584543}{1344} Y^{2}-\frac{1745175}{1792} Y^{3}\right) q^{3} \cr
&&
+\left(-\frac{33}{256}+\frac{207}{448} Y -\frac{153}{128} Y^{2}
-\frac{7}{64} Y^{3}-\frac{525}{256} Y^{4}+\frac{15}{32} Y^{5}\right) q^{5}, \cr
\tilde{\dot{Y}}_0 &=& 0, \quad
\tilde{\dot{Y}}_1 = 0, \quad
\tilde{\dot{Y}}_2 = 0, \cr
\tilde{\dot{Y}}_3 &=& 1, \quad
\tilde{\dot{Y}}_4 = -\frac{13 Y q}{244}, \quad
\tilde{\dot{Y}}_5 =
-\frac{10545}{1708}+\left(-\frac{99}{976}-\frac{45 Y^{2}}{976}\right) q^{2}, \cr
\tilde{\dot{Y}}_6 &=&
\frac{290 \pi}{61}+\left(-\frac{22}{3}+\frac{24821 Y}{3416}\right) q, \quad
\tilde{\dot{Y}}_7 = 
\frac{5177}{7686}-\frac{97 Y \pi  q}{122}+\left(-\frac{1477}{1952}
+\frac{723}{122} Y -\frac{4867}{1952} Y^{2}\right) q^{2}, \cr
\tilde{\dot{Y}}_8 &=&
\frac{6 Y \left(3+Y^{2}\right) \Psi_B^{(0,2)}(q)}{61}
-\frac{12 Y^{3}}{61} \Psi_B^{(0,1)}(q)
-\frac{22571 \pi}{854}
+\left(\frac{45695}{854}+\frac{408240 \kappa -20321879}{368928}Y\right) q \cr
&&
+\left[\frac{18 Y \left(3+Y^{2}\right) \Psi_B^{(0,2)}(q)}{61}
+\frac{9 Y^{3}}{61} \Psi_B^{(0,1)}(q) \right] q^{2} \cr
&&
+\left(\frac{429}{488}+\frac{720 \kappa -5435}{1952} Y
+\frac{39}{488} Y^2+\frac{720 \kappa +1811}{1952} Y^3 \right) q^{3}, \cr
\tilde{\dot{Y}}_9 &=&
-\frac{130796 \ln x}{6405}+\frac{1348 \pi^{2}}{183}
-\frac{261164 \ln \! \left(2\right)}{6405}-\frac{130796 \gamma}{6405}
+\frac{98625105067}{710186400}+\left(-\frac{8120}{183} \pi 
+\frac{315059}{6832} \pi  Y \right) q \cr
&&
+\left(\frac{3626515}{70272}-\frac{1557155}{10248} Y
+\frac{68911}{896} Y^{2}\right) q^{2}
+\left(-\frac{7281}{27328}-\frac{1287}{1952} Y +\frac{405}{488} Y^{2}
-\frac{585}{1952} Y^{3}+\frac{735}{3904} Y^{4}\right) q^{4}. \cr
\end{eqnarray*}

\section{Expansion coefficients in Eq.~\eqref{eq:Yn}}
\label{sec:coeffs-Yn}

The expansion coefficients of $Y(x)$ in Eq.~\eqref{eq:Yn}, $\tilde{Y}_k$,
for $k\le 9$ are given by
\begin{eqnarray*}
\tilde{Y}_3 &=& {\frac{61}{72}}, \quad
\tilde{Y}_4 = -\frac{13 \tilde{Y}_0 q}{384}, \quad
\tilde{Y}_5 =
-\frac{81217}{40320} - \left( \frac{3 \tilde{Y}_0^{2}}{128} + \frac{33}{640}\right) q^{2}, \quad
\tilde{Y}_6 = \frac{23 \pi}{72}+\left(\frac{3671095 \tilde{Y}_0}{580608}-\frac{61}{36}\right) q, \cr
\tilde{Y}_7 &=&
-\frac{617846599}{170698752}-\frac{71 \pi  \tilde{Y}_0 q}{336} +\left(-\frac{2824223}{903168} \tilde{Y}_0^{2}
+\frac{49}{36} \tilde{Y}_0 +\frac{1467113}{2709504}\right) q^{2}, \cr
\tilde{Y}_8 &=&
-\frac{\tilde{Y}_0^{3}}{16} \Psi_B^{(0,1)}(q)
+\frac{\tilde{Y}_0 (\tilde{Y}_0^{2}+3)}{32} \Psi_B^{(0,2)}(q)
-\frac{111617 \pi}{129024}+\left(\frac{45}{128} \tilde{Y}_0 \kappa 
-\frac{23469357001}{3901685760} \tilde{Y}_0 +\frac{81217}{12096}\right) q \cr
&&
+\left[\frac{3 \tilde{Y}_0^{3}}{64} \Psi_B^{(0,1)}(q)
+\frac{3 \tilde{Y}_0 (\tilde{Y}_0^{2}+3)}{32} \Psi_B^{(0,2)}(q)
+\frac{15 \pi  \,\tilde{Y}_0^{2}}{256}+\frac{33 \pi}{256}\right] q^{2} \cr
&&
+\left[\frac{11}{64}
+\frac{\left(172800 \kappa +567235\right) \tilde{Y}_0^{3}}{1474560}+\frac{\tilde{Y}_0^{2}}{96}
+\frac{\left(172800 \kappa -1615377\right) \tilde{Y}_0}{1474560}\right] q^{3}, \cr
\tilde{Y}_9 &=&
-\frac{5 \pi^{2}}{18}-\frac{2197 \gamma}{1890}-\frac{2615 \ln \! \left(2\right)}{1134}
-\frac{2197 \ln x}{1890}+\frac{40135977646837}{20279011737600} \cr
&&
+\left(\frac{2335369}{580608} \pi  \tilde{Y}_0 -\frac{23}{18} \pi \right) q
+\left(\frac{255882808969}{5852528640} \tilde{Y}_0^{2}-\frac{3280}{81} \tilde{Y}_0 
+\frac{47697759001}{17557585920}\right) q^{2} \cr
&&+\left(\frac{4439}{36864} \tilde{Y}_0^{4}
-\frac{15}{256} \tilde{Y}_0^{3}+\frac{66139}{184320} \tilde{Y}_0^{2}-\frac{33}{256} \tilde{Y}_0 
-\frac{93893}{645120}\right) q^{4}.
\end{eqnarray*}

\section{Expansion coefficients in $t_{\textrm{T2}}^{(n)}$ and
$\Phi_{\textrm{T2}}^{(n)}$}
\label{sec:coeffs-tPhi_T2}

The expansion coefficients in Eqs.~\eqref{eq:t_T2} and~\eqref{eq:Phi_T2},
$\tilde{t}_k$ and $\tilde{\Phi}_k^{\textrm{T2}}$, for $k\le 9$
are given as
\begin{eqnarray*}
\tilde{t}_0 &=& 1, \quad
\tilde{t}_1 = 0, \quad
\tilde{t}_2 = {\frac{743}{252}}, \quad
\tilde{t}_3 =
-\frac{32 \pi}{5}+\left(\frac{16}{3}+\frac{146 \tilde{Y}_0}{15}\right) q, \cr
\tilde{t}_4 &=&
\frac{3058673}{508032} 
+ \left(\frac{233}{48}-4 \tilde{Y}_0 -\frac{527}{48} \tilde{Y}_0^{2}\right) q^{2}, \cr
\tilde{t}_5 &=& 
-\frac{7729 \pi}{252}+\left(\frac{743}{63}+\frac{138185 \tilde{Y}_0}{756}\right) q
+\left(\frac{3}{4} \tilde{Y}_0 +\frac{5}{4} \tilde{Y}_0^{3}\right) q^{3}, \cr
\tilde{t}_6 &=&
\frac{6848 \ln x}{105}-\frac{10052469856691}{23471078400}
+\frac{128 \pi^{2}}{3}+\frac{13696 \ln \! \left(2\right)}{105}+\frac{6848 \gamma}{105}
+\left(-\frac{64}{3} \pi -130 \pi  \tilde{Y}_0 \right) q \cr
&&
+\left(\frac{304897}{12096}+\frac{14011}{252} \tilde{Y}_0 -\frac{1611751}{12096} \tilde{Y}_0^{2}\right) q^{2}, \cr
\tilde{t}_7 &=& 
-\frac{15419335 \pi}{127008}+\left(\frac{3058673}{190512}
+\frac{5035793419 \tilde{Y}_0}{1524096}\right) q 
+\frac{\pi  (-153+32 \tilde{Y}_0 +447 \tilde{Y}_0^{2}) q^{2}}{2} \cr
&& +\left(\frac{49}{3}+\frac{2518475}{12096} \tilde{Y}_0 -105 \tilde{Y}_0^{2}-\frac{3756875}{12096} \tilde{Y}_0^{3}\right) q^{3}, \cr
\tilde{t}_8 &=& 
\Biggl\{-\frac{202204 \ln \! \left(2\right)}{441}-\frac{47385 \ln \! \left(3\right)}{196}
-\frac{18098 \pi^{2}}{63}-\frac{36812 \gamma}{105}+4 \kappa 
+\frac{2498644552172012833}{461347517030400} \cr
&&
+\left[(4-4 \tilde{Y}_0^{4}) \Psi_B^{(0,1)}(q)
+(2+12 \tilde{Y}_0^{2}+2 \tilde{Y}_0^{4}) \Psi_B^{(0,2)}(q)
+\frac{141245 \pi  \tilde{Y}_0}{56}\right] q \cr
&& +\left[7 \kappa -\frac{8416499209}{8709120}
+\frac{\left(2743372800 \kappa +36138488713\right) \tilde{Y}_0^{2}}{60963840}\right] q^{2} \cr
&&
+\left[(-3+3 \tilde{Y}_0^{4}) \Psi_B^{(0,1)}(q)
+(6+36 \tilde{Y}_0^{2}+6 \tilde{Y}_0^{4}) \Psi_B^{(0,2)}(q) 
+18 \pi  \tilde{Y}_0 +30 \pi  \,\tilde{Y}_0^{3}\right] q^{3} \cr
&&
+\left(\frac{3}{2} \kappa -\frac{14819}{5760}
+15 \kappa  \,\tilde{Y}_0^{2}+\frac{254401}{3840} \tilde{Y}_0^{2}+\frac{15}{2} \kappa  \,\tilde{Y}_0^{4}
-\frac{492385}{2304} \tilde{Y}_0^{4}\right) q^{4}\Biggr\} \ln x
-\frac{18406 (\ln x)^2}{105}, \cr
\tilde{t}_9 &=& 
\left[\frac{54784 \pi}{105}+\left(\frac{27392}{315}-\frac{363112 \tilde{Y}_0}{315}\right) q \right] 
\ln x + \frac{512 \pi}{3} \left(\pi^{2}+\frac{107 \gamma}{35}
+\frac{214 \ln \left(2\right)}{35}-\frac{102282756713483}{4005730713600}\right) \cr
&&
+\biggl[\frac{(3061522759680-40631789721600 \tilde{Y}_0 ) \ln \! \left(2\right)}{17603308800}
+\frac{512 \pi^{2}}{9}+\frac{27392 \gamma}{315}-\frac{10817850546611}{17603308800} \cr
&&
+\frac{\left(-15287495731200 \pi^{2}-20291976714240 \gamma
-153221817336999\right) \tilde{Y}_0}{17603308800}\biggr] q \cr
&&
-\frac{\pi  \left(-5542607+3088440 \tilde{Y}_0 +8488481 \tilde{Y}_0^{2}\right) q^{2}}{12096} \cr
&&
+\left(\frac{76391}{2160}+\frac{31187580643}{60963840} \tilde{Y}_0 -\frac{5064011}{45360} \tilde{Y}_0^{2}
+\frac{82597339637}{60963840} \tilde{Y}_0^{3}\right) q^{3} \cr
&&
+\left(-\frac{33}{160}
-\frac{49687}{17920} \tilde{Y}_0 -\frac{81}{80} \tilde{Y}_0^{2}+\frac{57}{20} \tilde{Y}_0^{3}
-\frac{57}{32} \tilde{Y}_0^{4}+\frac{22141}{512} \tilde{Y}_0^{5}\right) q^{5}, \cr
\tilde{\Phi}_0^{\textrm{T2}} &=& 1, \quad
\tilde{\Phi}_1^{\textrm{T2}} = 0, \quad
\tilde{\Phi}_2^{\textrm{T2}} = {\frac{3715}{1008}}, \quad
\tilde{\Phi}_3^{\textrm{T2}} =
-10 \pi +\left(\frac{25}{3}+\frac{365 \tilde{Y}_0}{24}\right) q, \cr
\tilde{\Phi}_4^{\textrm{T2}} &=&
\frac{15293365}{1016064}
+\left(\frac{1165}{96}-10 \tilde{Y}_0 -\frac{2635}{96} \tilde{Y}_0^{2}\right) q^{2}, \cr
\tilde{\Phi}_5^{\textrm{T2}} &=&
\left[\frac{38645 \pi}{672}+\left(-\frac{3715}{168}-\frac{690925 \tilde{Y}_0}{2016}\right) q
+\left(-\frac{45}{32} \tilde{Y}_0 -\frac{75}{32} \tilde{Y}_0^{3}\right) q^{3}\right] \ln x, \cr
\tilde{\Phi}_6^{\textrm{T2}} &=&
-\frac{1712 \ln x}{21}-\frac{160 \pi^{2}}{3}
-\frac{3424 \ln \! \left(2\right)}{21}-\frac{1712 \gamma}{21}
+\frac{12348611926451}{18776862720}
+\left(\frac{80}{3} \pi +\frac{325}{2} \pi  \tilde{Y}_0 \right) q \cr
&&
+\left(-\frac{1524485}{48384}-\frac{70055}{1008} \tilde{Y}_0
+\frac{8058755}{48384} \tilde{Y}_0^{2}\right) q^{2}, \cr
\tilde{\Phi}_7^{\textrm{T2}} &=&
\frac{77096675 \pi}{2032128}+\left(-\frac{15293365}{3048192}
-\frac{25178967095 \tilde{Y}_0}{24385536}\right) q
-\frac{5 \pi  \left(-153+32 \tilde{Y}_0 +447 \tilde{Y}_0^{2}\right) q^{2}}{32} \cr
&&
+\left(-\frac{245}{48}-\frac{12592375}{193536} \tilde{Y}_0 +\frac{525}{16} \tilde{Y}_0^{2}
+\frac{18784375}{193536} \tilde{Y}_0^{3}\right) q^{3}, \cr
\tilde{\Phi}_8^{\textrm{T2}} &=&
-\frac{9203 \ln x}{126}-\frac{252755 \ln \! \left(2\right)}{2646}
-\frac{78975 \ln \! \left(3\right)}{1568}-\frac{45245 \pi^{2}}{756}
-\frac{9203 \gamma}{126}+\frac{5 \kappa}{6}+\frac{2552559234067006753}{2214468081745920} \cr
&&
+\left[\left(\frac{5}{6}-\frac{5 \tilde{Y}_0^{4}}{6}\right) \Psi_B^{(0,1)}(q)
+\left(\frac{5}{12}+\frac{5}{2} \tilde{Y}_0^{2}+\frac{5}{12} \tilde{Y}_0^{4}\right) \Psi_B^{(0,2)}(q)
+\frac{706225 \pi  \tilde{Y}_0}{1344}\right] q \cr
&&
+\left(\frac{35 \kappa}{24}-\frac{8416499209}{41803776}
+\frac{\left(2743372800 \kappa +36138488713\right) \tilde{Y}_0^{2}}{292626432}\right) q^{2} \cr
&&
+\left[\left(-\frac{5}{8}+\frac{5 \tilde{Y}_0^{4}}{8}\right) \Psi_B^{(0,1)}(q)
+\left(\frac{5}{4}+\frac{15}{2} \tilde{Y}_0^{2}+\frac{5}{4} \tilde{Y}_0^{4}\right) \Psi_B^{(0,2)}(q)
+\frac{5 \pi (3+5 \tilde{Y}_0^{2}) \tilde{Y}_0}{4}\right] q^{3} \cr
&&
+\left[\frac{5 \kappa}{16}
-\frac{14819}{27648}+\frac{\left(172800 \kappa +763203\right) \tilde{Y}_0^{2}}{55296}
+\frac{\left(86400 \kappa -2461925\right) \tilde{Y}_0^{4}}{55296}\right] q^{4}, \cr
\tilde{\Phi}_9^{\textrm{T2}} &=&
\left[\frac{1712 \pi}{21}+\left(\frac{856}{63}-\frac{45389 \tilde{Y}_0}{252}\right) q \right]
\ln x
+\frac{80 \pi}{3} \left(\pi^{2}+\frac{107 \gamma}{35}+\frac{214 \ln \left(2\right)}{35}
-\frac{93098188434443}{4005730713600}\right) \cr
&&
+ \biggl[\frac{(3061522759680-40631789721600 \tilde{Y}_0 ) \ln \! \left(2\right)}{112661176320}
+\frac{80 \pi^{2}}{9}+\frac{856 \gamma}{63}-\frac{9669779511731}{112661176320} \cr
&&
+\frac{\left(-15287495731200 \pi^{2}-20291976714240 \gamma 
-168440799872679\right) \tilde{Y}_0}{112661176320}\biggr] q \cr
&&
-\frac{5 \pi (-5542607+3088440 \tilde{Y}_0 +8488481 \tilde{Y}_0^{2}) q^{2}}{387072} \cr
&&
+\left(\frac{76391}{13824}+\frac{31187580643}{390168576} \tilde{Y}_0 -\frac{5064011}{290304} \tilde{Y}_0^{2}
+\frac{82597339637}{390168576} \tilde{Y}_0^{3}\right) q^{3} \cr
&&
+\left(-\frac{33}{1024}
-\frac{49687}{114688} \tilde{Y}_0 -\frac{81}{512} \tilde{Y}_0^{2}+\frac{57}{128} \tilde{Y}_0^{3}
-\frac{285}{1024} \tilde{Y}_0^{4}+\frac{110705}{16384} \tilde{Y}_0^{5}\right) q^{5}.
\end{eqnarray*}

\section{Expansion coefficients in $\Phi_{\textrm{T3}}^{(n)}$}
\label{sec:coeffs-Phi_T3}

The expansion coefficients in Eq.~\eqref{eq:Phi_T3},
$\tilde{\Phi}_k^{\textrm{T3}}$, for $k\le 9$
are given as
\begin{eqnarray*}
\tilde{\Phi}_0^{\textrm{T3}} &=& 1, \quad
\tilde{\Phi}_1^{\textrm{T3}} = 0, \quad
\tilde{\Phi}_2^{\textrm{T3}} = {\frac{3715}{8064}}, \quad
\tilde{\Phi}_3^{\textrm{T3}} =
-\frac{3 \pi}{4}+\left(\frac{5}{8}+\frac{73 \tilde{Y}_0}{64}\right) q, \cr
\tilde{\Phi}_4^{\textrm{T3}} &=&
\frac{9275495}{14450688}+\left(\frac{1165}{2048}-\frac{15}{32} \tilde{Y}_0
-\frac{2635}{2048} \tilde{Y}_0^{2}\right) q^{2}, \cr
\tilde{\Phi}_5^{\textrm{T3}} &=&
\left[\frac{38645 \pi}{21504}+\left(-\frac{3715}{5376}
-\frac{690925 \tilde{Y}_0}{64512}\right) q
+\left(-\frac{45}{1024} \tilde{Y}_0 -\frac{75}{1024} \tilde{Y}_0^{3}\right) q^{3}\right]
\ln\frac{\Theta}{2} \cr
&&
+\frac{47561 \pi}{64512}+\left(-\frac{3715}{10752}-\frac{2926417 \tilde{Y}_0}{774144}\right) q
+\left(-\frac{15}{1024} \tilde{Y}_0 -\frac{25}{1024} \tilde{Y}_0^{3}\right) q^{3}, \cr
\tilde{\Phi}_6^{\textrm{T3}} &=& 
-\frac{107}{56} \ln\frac{\Theta}{2} - \frac{53 \pi^{2}}{40}
-\frac{107 \ln \! \left(2\right)}{28}-\frac{107 \gamma}{56}+\frac{831032450749357}{57682522275840}
+\left(\frac{3}{4} \pi +\frac{5167}{1280} \pi  \tilde{Y}_0 \right) q \cr
&&
+\left(-\frac{662935}{786432}-\frac{21821}{12288} \tilde{Y}_0 
+\frac{105938291}{27525120} \tilde{Y}_0^{2}\right) q^{2}, \cr
\tilde{\Phi}_7^{\textrm{T3}} &=&
\frac{188516689 \pi}{173408256}+\left(-\frac{3008935}{14450688}
-\frac{6564218461 \tilde{Y}_0}{260112384}\right) q 
-\frac{\pi (-79+21 \tilde{Y}_0 +226 \tilde{Y}_0^{2}) q^{2}}{128} \cr
&&
+\left(-\frac{4105}{24576}-\frac{3254407}{2064384} \tilde{Y}_0
+\frac{23287}{24576} \tilde{Y}_0^{2}+\frac{1271527}{516096} \tilde{Y}_0^{3}\right) q^{3}, \cr
\tilde{\Phi}_8^{\textrm{T3}} &=&
\frac{9203}{21504} \left(\ln\frac{\Theta}{2}\right)^{2}
+\Biggl\{\frac{252755 \ln \! \left(2\right)}{225792}
+\frac{236925 \ln \! \left(3\right)}{401408}+\frac{45245 \pi^{2}}{64512}
+\frac{9203 \gamma}{10752}-\frac{5 \kappa}{512}-\frac{2585826888444724513}{188967942975651840} \cr
&&
+\left[\left(-\frac{5}{512}+\frac{5 \tilde{Y}_0^{4}}{512}\right) \Psi_B^{(0,1)}(q)
+\left(-\frac{5}{1024}-\frac{15}{512} \tilde{Y}_0^{2}-\frac{5}{1024} \tilde{Y}_0^{4}\right) \Psi_B^{(0,2)}(q)
-\frac{706225 \pi  \tilde{Y}_0}{114688}\right] q \cr
&&
+\left[-\frac{35 \kappa}{2048}
+\frac{8416499209}{3567255552}+\frac{\left(-2743372800 \kappa
-36138488713\right) \tilde{Y}_0^{2}}{24970788864}\right] q^{2} \cr
&&
+\left[\left(\frac{15}{2048}-\frac{15 \tilde{Y}_0^{4}}{2048}\right) \Psi_B^{(0,1)}(q)
+\left(-\frac{15}{1024}-\frac{45}{512} \tilde{Y}_0^{2}-\frac{15}{1024} \tilde{Y}_0^{4}\right) \Psi_B^{(0,2)}(q)
-\frac{15 \pi  (3+5 \tilde{Y}_0^{2}) \tilde{Y}_0}{1024}\right] q^{3} \cr
&&
+\left[-\frac{15 \kappa}{4096}
+\frac{14819}{2359296}+\frac{\left(-172800 \kappa -763203\right) \tilde{Y}_0^{2}}{4718592}
+\frac{\left(-86400 \kappa +2461925\right) \tilde{Y}_0^{4}}{4718592}\right] q^{4}\Biggr\}
\ln\frac{\Theta}{2} \cr
&&
-\frac{245629 \ln \! \left(2\right)}{338688}-\frac{78975 \ln \! \left(3\right)}{401408}
-\frac{191257 \pi^{2}}{387072}-\frac{208343 \gamma}{451584}+\frac{5 \kappa}{1536}
+\frac{11722551188833191865}{2073248288647151616} \cr
&&
+\left[\left(\frac{5}{1536}
-\frac{5 \tilde{Y}_0^{4}}{1536}\right) \Psi_B^{(0,1)}(q)
+\left(\frac{5}{3072}+\frac{5}{512} \tilde{Y}_0^{2}+\frac{5}{3072} \tilde{Y}_0^{4}\right) \Psi_B^{(0,2)}(q)
+\frac{56477 \pi}{258048}+\frac{1879999 \pi  \tilde{Y}_0}{516096}\right] q \cr
&&
+\left(\frac{35}{6144} \kappa -\frac{2164129184801}{2397195730944}
-\frac{161539403}{154140672} \tilde{Y}_0 +\frac{75}{2048} \kappa  \,\tilde{Y}_0^{2}
-\frac{1614749247721}{2397195730944} \tilde{Y}_0^{2}\right) q^{2} \cr
&&
+\left[\left(-\frac{5}{2048}+\frac{5 \tilde{Y}_0^{4}}{2048}\right) \Psi_B^{(0,1)}(q)
+\left(\frac{5}{1024}+\frac{15}{512} \tilde{Y}_0^{2}+\frac{5}{1024} \tilde{Y}_0^{4}\right) \Psi_B^{(0,2)}(q)
+\frac{13 \pi (3+5 \tilde{Y}_0^{2}) \tilde{Y}_0}{2048}\right] q^{3} \cr
&&
+\left[\frac{5 \kappa}{4096}-\frac{2917213}{226492416}+\frac{925 \tilde{Y}_0}{65536}
+\frac{\left(2764800 \kappa +20089830\right) \tilde{Y}_0^{2}}{226492416} -\frac{3035 \tilde{Y}_0^{3}}{65536}
+\frac{\left(1382400 \kappa -54411485\right) \tilde{Y}_0^{4}}{226492416}\right] q^{4}, \cr
\tilde{\Phi}_9^{\textrm{T3}} &=& 
\left[-\frac{321 \pi}{1120}+\left(-\frac{107}{448}+\frac{54819 \tilde{Y}_0}{71680}\right) q \right]
\ln\frac{\Theta}{2} 
-\frac{33 \pi}{800}  \left(\pi^{2}+\frac{535 \gamma}{77}+\frac{1070 \ln \left(2\right)}{77}
-\frac{587519428177201}{7931346812928}\right) \cr
&&
+\Biggl[\frac{(-137768524185600+441858474731520 \tilde{Y}_0 )
\ln \! \left(2\right)}{288412611379200}-\frac{31 \pi^{2}}{160}-\frac{107 \gamma}{448} \cr
&&
+\frac{46133573278267}{28841261137920}
+\frac{\left(77798175307776 \pi^{2}+220570465167360 \gamma +1400875984590511\right) \tilde{Y}_0}
{288412611379200}\Biggr] q \cr
&&
+\frac{\pi (-180539475+740494160 \tilde{Y}_0 +389275781 \tilde{Y}_0^{2}) q^{2}}{1101004800} \cr
&&
+\left(-\frac{11782937}{99090432}-\frac{11104800985}{8323596288} \tilde{Y}_0
+\frac{350534719}{495452160} \tilde{Y}_0^{2} 
+\frac{135670708597}{208089907200} \tilde{Y}_0^{3}\right) q^{3} \cr
&&
+\left(\frac{99}{524288}+\frac{51201}{58720256} \tilde{Y}_0 +\frac{603}{262144} \tilde{Y}_0^{2}
-\frac{53}{32768} \tilde{Y}_0^{3}+\frac{2055}{524288} \tilde{Y}_0^{4}
-\frac{279415}{8388608} \tilde{Y}_0^{5}\right) q^{5}.
\end{eqnarray*}

\section{Expansion coefficients in $\psi_{\textrm{F2}}^{(n)}$}
\label{sec:coeffs-psi_F2}

The expansion coefficients in Eq.~\eqref{eq:psi_F2},
$\tilde{\psi}_k^{\textrm{F2}}$ for $k\le 9$ are given as
\begin{eqnarray*}
\tilde{\psi}_0^{\textrm{F2}} &=& 1, \quad
\tilde{\psi}_1^{\textrm{F2}} = 0, \quad
\tilde{\psi}_2^{\textrm{F2}} = {\frac{3715}{756}}, \quad
\tilde{\psi}_3^{\textrm{F2}} =
-16 \pi +\left(\frac{40}{3}+\frac{73 \tilde{Y}_0}{3}\right) q, \cr
\tilde{\psi}_4^{\textrm{F2}} &=& 
\frac{15293365}{508032}+\left(\frac{1165}{48}-20 \tilde{Y}_0 -\frac{2635}{48} \tilde{Y}_0^{2}\right) q^{2}, \cr
\tilde{\psi}_5^{\textrm{F2}} &=& 
\left[\left(-\frac{15}{4} \tilde{Y}_0 -\frac{25}{4} \tilde{Y}_0^{3}\right) q^{3}
+\left(-\frac{3715}{63}-\frac{690925 \tilde{Y}_0}{756}\right) q
+\frac{38645 \pi}{252}\right] \ln x_f, \cr
\tilde{\psi}_6^{\textrm{F2}} &=&
-\frac{6848 \ln x_f}{21}-\frac{640 \pi^{2}}{3}
-\frac{6848 \gamma}{21}-\frac{13696 \ln \! \left(2\right)}{21}
+\frac{11583231236531}{4694215680} \cr
&&
+\left(\frac{320}{3} \pi +650 \pi  \tilde{Y}_0 \right) q
+\left(-\frac{1524485}{12096}-\frac{70055}{252} \tilde{Y}_0
+\frac{8058755}{12096} \tilde{Y}_0^{2}\right) q^{2}, \cr
\tilde{\psi}_7^{\textrm{F2}} &=&
\frac{77096675 \pi}{254016}
+\left(-\frac{15293365}{381024}-\frac{25178967095 \tilde{Y}_0}{3048192}\right) q
-\frac{5 \pi (-153+32 \tilde{Y}_0 +447 \tilde{Y}_0^{2}) q^{2}}{4} \cr
&&
+\left(-\frac{245}{6}-\frac{12592375}{24192} \tilde{Y}_0 +\frac{525}{2} \tilde{Y}_0^{2}
+\frac{18784375}{24192} \tilde{Y}_0^{3}\right) q^{3}, \cr
\tilde{\psi}_8^{\textrm{F2}} &=&
-\frac{90490 \pi^{2}}{567}-\frac{26325 \ln \! \left(3\right)}{196}
-\frac{36812 \gamma}{189}-\frac{1011020 \ln \! \left(2\right)}{3969}
+\frac{20 \kappa}{9}+\frac{2552559234067006753}{830425530654720} \cr
&&
+\left[\left(\frac{20}{9}-\frac{20 \tilde{Y}_0^{4}}{9}\right) \Psi_B^{(0,1)}(q)
+\left(\frac{10}{9}+\frac{20}{3} \tilde{Y}_0^{2}+\frac{10}{9} \tilde{Y}_0^{4}\right) \Psi_B^{(0,2)}(q)
+\frac{706225 \pi  \tilde{Y}_0}{504}\right] q \cr
&&
+\left(\frac{35 \kappa}{9}-\frac{8416499209}{15676416}
+\frac{\left(2743372800 \kappa +36138488713\right) \tilde{Y}_0^{2}}{109734912}\right) q^{2} \cr
&&
+\left[\left(-\frac{5}{3}+\frac{5 \tilde{Y}_0^{4}}{3}\right) \Psi_B^{(0,1)}(q)
+\left(\frac{10}{3}+20 \tilde{Y}_0^{2}+\frac{10}{3} \tilde{Y}_0^{4}\right) \Psi_B^{(0,2)}(q)
+10 \pi  \tilde{Y}_0 +\frac{50 \pi  \,\tilde{Y}_0^{3}}{3}\right] q^{3} \cr
&&
+\left[\frac{5 \kappa}{6}-\frac{14819}{10368}+\frac{\left(172800 \kappa +763203\right) \tilde{Y}_0^{2}}{20736}+\frac{\left(86400 \kappa -2461925\right) \tilde{Y}_0^{4}}{20736}\right] q^{4} \cr
&&
+\Biggl\{\frac{90490 \pi^{2}}{189}
+\frac{78975 \ln \! \left(3\right)}{196}+\frac{36812 \gamma}{63}
+\frac{1011020 \ln \! \left(2\right)}{1323}-\frac{20 \kappa}{3}
-\frac{2552559234067006753}{276808510218240} \cr
&&
+\left[\left(-\frac{20}{3}+\frac{20 \tilde{Y}_0^{4}}{3}\right) \Psi_B^{(0,1)}(q)
+\left(-\frac{10}{3}-20 \tilde{Y}_0^{2}-\frac{10}{3} \tilde{Y}_0^{4}\right) \Psi_B^{(0,2)}(q)
-\frac{706225 \pi  \tilde{Y}_0}{168}\right] q \cr
&&
+\left[-\frac{35 \kappa}{3}+\frac{8416499209}{5225472}
+\frac{\left(-2743372800 \kappa -36138488713\right) \tilde{Y}_0^{2}}{36578304}\right] q^{2} \cr
&&
+\left[(5-5 \tilde{Y}_0^{4}) \Psi_B^{(0,1)}(q)
+(-10-60 \tilde{Y}_0^{2}-10 \tilde{Y}_0^{4}) \Psi_B^{(0,2)}(q)
-30 \pi  \tilde{Y}_0 -50 \pi  \,\tilde{Y}_0^{3}\right] q^{3} \cr
&&
+\left[-\frac{5 \kappa}{2}+\frac{14819}{3456}
+\frac{\left(-172800 \kappa -763203\right) \tilde{Y}_0^{2}}{6912}
+\frac{\left(-86400 \kappa +2461925\right) \tilde{Y}_0^{4}}{6912}\right] q^{4}
\Biggr\} \ln x_f \cr
&&
+ \frac{18406}{63} (\ln x_f)^2, \cr
\tilde{\psi}_9^{\textrm{F2}} &=&
-\frac{640 \pi}{3}  \left(\pi^{2}+\frac{107 \gamma}{35}+\frac{214 \ln \left(2\right)}{35}
-\frac{105344279473163}{4005730713600}\right) \cr
&&
+\Biggl[\frac{(-3061522759680+40631789721600 \tilde{Y}_0 ) \ln \! \left(2\right)}{14082647040}
-\frac{640 \pi^{2}}{9}-\frac{6848 \gamma}{63}+\frac{11200540891571}{14082647040} \cr
&&
+\frac{\left(15287495731200 \pi^{2}+20291976714240 \gamma
+148148823158439\right) \tilde{Y}_0}{14082647040}\Biggr] q \cr
&&
+\frac{5 \pi  \left(-5542607+3088440 \tilde{Y}_0 +8488481 \tilde{Y}_0^{2}\right) q^{2}}{48384} \cr
&&
+\left(-\frac{76391}{1728}-\frac{31187580643}{48771072} \tilde{Y}_0 
+\frac{5064011}{36288} \tilde{Y}_0^{2}-\frac{82597339637}{48771072} \tilde{Y}_0^{3}\right) q^{3} \cr
&&
+\left(\frac{33}{128}+\frac{49687}{14336} \tilde{Y}_0 +\frac{81}{64} \tilde{Y}_0^{2}
-\frac{57}{16} \tilde{Y}_0^{3}+\frac{285}{128} \tilde{Y}_0^{4}-\frac{110705}{2048} \tilde{Y}_0^{5}\right) q^{5} \cr
&&
+ \left[-\frac{13696 \pi}{21}+\left(-\frac{6848}{63}+\frac{90778 \tilde{Y}_0}{63}\right) q \right] \ln x_f.
\end{eqnarray*}

\section{Expansion coefficients of $\mathcal{E}_n(x)$ and $\mathcal{F}_n(x)$}
\label{sec:coeffs-calEF}

The expansion coefficients in Eqs.~\eqref{eq:calEn} and~\eqref{eq:calFn},
$\tilde{\mathcal{E}}_k$ and $\tilde{\mathcal{F}}_k$, for $k\le 9$
are given as
\begin{eqnarray*}
\tilde{\mathcal{E}}_0 &=& 1, \quad
\tilde{\mathcal{E}}_1 = 0, \quad
\tilde{\mathcal{E}}_2 = -{\frac{3}{4}}, \quad
\tilde{\mathcal{E}}_3 = \left(-\frac{4}{3}+4 \tilde{Y}_0 \right) q, \quad
\tilde{\mathcal{E}}_4 =
-\frac{27}{8}+\left(\frac{1}{2}+\tilde{Y}_0 -\frac{5}{2} \tilde{Y}_0^{2}\right) q^{2}, \cr
\tilde{\mathcal{E}}_5 &=&
\left(2+6 \tilde{Y}_0 \right) q, \quad
\tilde{\mathcal{E}}_6 =
-\frac{675}{64}+\left(\frac{73}{12}-\frac{137}{6} \tilde{Y}_0
+\frac{473}{36} \tilde{Y}_0^{2}\right) q^{2}, \cr
\tilde{\mathcal{E}}_7 &=&
\left(-\frac{205}{72}-\frac{1333}{288} \tilde{Y}_0 +\frac{2365}{72} \tilde{Y}_0^{2}
-\frac{7307}{288} \tilde{Y}_0^{3}\right) q^{3}+\left(\frac{27}{2}
+\frac{27 \tilde{Y}_0}{2}\right) q, \cr
\tilde{\mathcal{E}}_8 &=& 
-\frac{3969}{128}+\left(\frac{69}{80}+\frac{925}{384} \tilde{Y}_0 
-\frac{1621}{1920} \tilde{Y}_0^{2}-\frac{4861}{384} \tilde{Y}_0^{3}
+\frac{3929}{384} \tilde{Y}_0^{4}\right) q^{4}+\left(\frac{52867}{10080}
-\frac{209}{8} \tilde{Y}_0 +\frac{13493}{10080} \tilde{Y}_0^{2}\right) q^{2}, \cr
\tilde{\mathcal{E}}_9 &=&
\frac{3 \left(11+5 \tilde{Y}_0^{2}\right) \left(1-5 \tilde{Y}_0 \right)
\left(1-\tilde{Y}_0^{2}\right) q^{5}}{640}
-\left(\frac{8641487}{362880}
-\frac{17133971}{145152} \tilde{Y}_0 + \frac{25781051}{120960} \tilde{Y}_0^{2}
-\frac{17976685}{145152} \tilde{Y}_0^{3}\right) q^{3} \cr
&&
+\left(-\frac{23}{18} \pi +\frac{23}{18} \pi  \,\tilde{Y}_0^{2}\right) q^{2}
+\left(\frac{225}{4}+\frac{135 \tilde{Y}_0}{4}\right) q, \cr
\tilde{\mathcal{F}}_0 &=& 1, \quad
\tilde{\mathcal{F}}_1 = 0, \quad
\tilde{\mathcal{F}}_2 = -{\frac{1247}{336}}, \quad
\tilde{\mathcal{F}}_3 = 4 \pi +\left(-\frac{20}{3}+\frac{47 \tilde{Y}_0}{12}\right) q, \cr
\tilde{\mathcal{F}}_4 &=&
 -\frac{44711}{9072}+\left(-\frac{89}{96}+5 \tilde{Y}_0
 -\frac{193}{96} \tilde{Y}_0^{2}\right) q^{2}, \cr
\tilde{\mathcal{F}}_5 &=& 
-\frac{8191 \pi}{672}+\left(\frac{1247}{42}-\frac{11299 \tilde{Y}_0}{336}\right) q 
+\left(-\frac{9}{32} \tilde{Y}_0 -\frac{15}{32} \tilde{Y}_0^{3}\right) q^{3}, \cr
\tilde{\mathcal{F}}_6 &=&
 \frac{6643739519}{69854400}+\frac{16 \pi^{2}}{3}-\frac{1712 \gamma}{105}
-\frac{3424 \ln \! \left(2\right)}{105}+\left(-\frac{104}{3} \pi
 +\frac{143}{6} \pi  \tilde{Y}_0 \right) q -\frac{1712 \ln x}{105} \cr
&&
+\left(\frac{310333}{6048}-\frac{48491}{504} \tilde{Y}_0 
+\frac{278891}{6048} \tilde{Y}_0^{2}\right) q^{2}, \cr
\tilde{\mathcal{F}}_7 &=&
 -\frac{16285 \pi}{504}+\left(\frac{44711}{972}-\frac{6899 \tilde{Y}_0}{1296}\right) q
 -\frac{\pi  \left(185-1248 \tilde{Y}_0 +673 \tilde{Y}_0^{2}\right) q^{2}}{48} \cr
&&
+\left(\frac{53}{12}-\frac{1092851}{13824} \tilde{Y}_0 +\frac{2603}{24} \tilde{Y}_0^{2}
-\frac{536941}{13824} \tilde{Y}_0^{3}\right) q^{3}, \cr
\tilde{\mathcal{F}}_8 &=&
 \frac{39931 \ln \! \left(2\right)}{294}-\frac{47385 \ln \! \left(3\right)}{1568}
-\frac{1369 \pi^{2}}{126}+\frac{232597 \gamma}{4410}+\frac{\kappa}{2}
-\frac{321516361867}{3178375200} \cr
&&
+\left[\left(\frac{1}{2}
-\frac{\tilde{Y}_0^{4}}{2}\right) \Psi_B^{(0,1)}(q)
+\left(\frac{1}{4}+\frac{3}{2} \tilde{Y}_0^{2}+\frac{1}{4} \tilde{Y}_0^{4}\right)
 \Psi_B^{(0,2)}(q)
+\frac{40955 \pi}{336}-\frac{49571 \pi  \tilde{Y}_0}{336}\right] q 
+\frac{232597 \ln x}{4410} \cr
&&
+\left[\frac{7 \kappa}{8}
-\frac{347474543}{1451520}+\frac{4353865 \tilde{Y}_0}{9072}
+\frac{\left(8164800 \kappa -333423217\right) \tilde{Y}_0^{2}}{1451520}\right] q^{2} \cr
&&
+\left[\left(-\frac{3}{8}+\frac{3 \tilde{Y}_0^{4}}{8}\right) \Psi_B^{(0,1)}(q)
+\left(\frac{3}{4}+\frac{9}{2} \tilde{Y}_0^{2}
+\frac{3}{4} \tilde{Y}_0^{4}\right) \Psi_B^{(0,2)}(q)\right] q^{3} \cr
&&
+\left[\frac{3 \kappa}{16}+\frac{15977}{2560}+\frac{111 \tilde{Y}_0}{128}
+\frac{\left(172800 \kappa +1332583\right) \tilde{Y}_0^{2}}{92160}
-\frac{4503 \tilde{Y}_0^{3}}{128}+\frac{\left(86400 \kappa +1329365\right) 
\tilde{Y}_0^{4}}{92160}\right] q^{4}, \cr
\tilde{\mathcal{F}}_9 &=& 
\left[-\frac{6848 \pi}{105}+\left(\frac{54784}{315}
-\frac{34261 \tilde{Y}_0}{315}\right) q \right] \ln x
-\frac{6848 \pi}{105}  \left(\gamma +2 \ln \! \left(2\right)
-\frac{265978667519}{48595599360}\right) \cr
&&
+\biggl[\frac{(18223349760-11414375280 \tilde{Y}_0 ) \ln \! \left(2\right)}{52390800}
-\frac{512 \pi^{2}}{9}+\frac{54784 \gamma}{315}-\frac{6572554559}{6548850} \cr
&&
+\frac{\left(2229519600 \pi^2-5698289520 \gamma +42430399563\right) \tilde{Y}_0}{52390800}\biggr] q
+\frac{(3655943-6696825 \tilde{Y}_0 +3192928 \tilde{Y}_0^{2}) \pi  \,q^{2}}{12096} \cr
&&
+\left(-\frac{23640385}{72576}+\frac{43398950927}{34836480} \tilde{Y}_0 
-\frac{16151149}{10368} \tilde{Y}_0^{2}+\frac{20873726833}{34836480} \tilde{Y}_0^3 \right) q^3 \cr
&&
+\left(\frac{33}{128}-\frac{117727}{143360} \tilde{Y}_0 -\frac{9}{4} \tilde{Y}_0^2
+\frac{1103}{640} \tilde{Y}_0^{3}-\frac{465}{128} \tilde{Y}_0^{4}
+\frac{18101}{4096} \tilde{Y}_0^{5}\right) q^{5}.
\end{eqnarray*}

\section{Expansion coefficients in Eq.~\eqref{eq:dotx_T4a}}
\label{sec:coeffs-dotx_T4a}

The expansion coefficients in Eq.~\eqref{eq:dotx_T4a},
$\tilde{\dot{x}}_k^{\textrm{T4a}}$, for $k\le 9$
are given as
\begin{eqnarray*}
\tilde{\dot{x}}_0^{\textrm{T4a}} &=& 1, \quad
\tilde{\dot{x}}_1^{\textrm{T4a}} = 0, \quad
\tilde{\dot{x}}_2^{\textrm{T4a}} = -{\frac{743}{336}}, \quad
\tilde{\dot{x}}_3^{\textrm{T4a}} = 
4 \pi +\left(-\frac{10}{3}-\frac{73 \tilde{Y}_0}{12}\right) q, \cr
\tilde{\dot{x}}_4^{\textrm{T4a}} &=&
\frac{34103}{18144}+\left(-\frac{233}{96}+2 \tilde{Y}_0 
+\frac{527}{96} \tilde{Y}_0^{2}\right) q^{2}, \cr
\tilde{\dot{x}}_5^{\textrm{T4a}} &=& 
-\frac{4159 \pi}{672}+\left(\frac{743}{72}-\frac{13991 \tilde{Y}_0}{336}\right) q
+\left(-\frac{9}{32} \tilde{Y}_0 -\frac{15}{32} \tilde{Y}_0^{3}\right) q^{3}, \cr
\tilde{\dot{x}}_6^{\textrm{T4a}} &=&
\frac{16 \pi^{2}}{3}-\frac{3424 \ln \! \left(2\right)}{105}-\frac{1712 \gamma}{105}
+\frac{16447322263}{139708800}+\left(-\frac{64}{3} \pi 
-\frac{97}{6} \pi  \tilde{Y}_0 \right) q -\frac{1712 \ln \! \left(x \right)}{105} \cr
&&
+\left(\frac{188015}{12096}+\frac{17953}{1008} \tilde{Y}_0 +\frac{556903}{12096} \tilde{Y}_0^{2}\right) q^{2}, \cr
\tilde{\dot{x}}_7^{\textrm{T4a}} &=&
-\frac{4415 \pi}{4032}+\left(-\frac{34103}{3024}-\frac{54595 \tilde{Y}_0}{336}\right) q
+\frac{\pi  \left(-473+672 \tilde{Y}_0 +767 \tilde{Y}_0^{2}\right) q^{2}}{48} \cr
&&
+\left(\frac{509}{36}-\frac{118691}{13824} \tilde{Y}_0 -\frac{1721}{36} \tilde{Y}_0^{2}-\frac{357949}{13824} \tilde{Y}_0^{3}\right) q^{3}, \cr
\tilde{\dot{x}}_8^{\textrm{T4a}} &=&
\frac{127751 \ln \! \left(2\right)}{1470}-\frac{47385 \ln \! \left(3\right)}{1568}
-\frac{361 \pi^{2}}{126}+\frac{124741 \gamma}{4410}+\frac{\kappa}{2}
+\frac{3971984677513}{25427001600} \cr
&&
+\left[\left(\frac{1}{2}-\frac{\tilde{Y}_0^{4}}{2}\right) \Psi_B^{(0,1)}(q)
+\left(\frac{1}{4}+\frac{3}{2} \tilde{Y}_0^{2}+\frac{1}{4} \tilde{Y}_0^{4}\right) \Psi_B^{(0,2)}(q)
+\frac{20795 \pi}{504}-\frac{5429 \pi  \tilde{Y}_0}{28}\right] q \cr
&&
+\frac{124741 \ln \! \left(x \right)}{4410}+\left(\frac{7}{8} \kappa 
-\frac{229555853}{1451520}+\frac{207683}{648} \tilde{Y}_0 +\frac{45}{8} \kappa  \,\tilde{Y}_0^{2}
+\frac{116682599}{207360} \tilde{Y}_0^{2}\right) q^{2} \cr
&&
+\left[\left(-\frac{3}{8}+\frac{3 \tilde{Y}_0^{4}}{8}\right) \Psi_B^{(0,1)}(q)
+\left(\frac{3}{4}+\frac{9}{2} \tilde{Y}_0^{2}+\frac{3}{4} \tilde{Y}_0^{4}\right)
\Psi_B^{(0,2)}(q) \right] q^{3} \cr
&&
+\left[\frac{3 \kappa}{16}+\frac{14257}{2560}-\frac{47 \tilde{Y}_0}{6}
+\frac{\left(172800 \kappa -1008617\right) \tilde{Y}_0^{2}}{92160}+\frac{301 \tilde{Y}_0^{3}}{12}
+\frac{\left(86400 \kappa +840965\right) \tilde{Y}_0^{4}}{92160}\right] q^{4}, \cr
\tilde{\dot{x}}_9^{\textrm{T4a}} &=&
-\frac{6848 \pi}{105}  \left(2 \ln \! \left(2\right)+\gamma -\frac{343801320119}{48595599360}\right) \cr
&&
+\biggl[\frac{(100228423680+45360120960 \tilde{Y}_0 ) \ln \! \left(2\right)}{419126400}
-\frac{352 \pi^{2}}{9}+\frac{37664 \gamma}{315}-\frac{16240238743}{19051200} \cr
&&
+\frac{\left(-4517251200 \pi^{2}+22751245440 \gamma -578438815461\right) \tilde{Y}_0}{419126400}
\biggr] q \cr
&&
+\left[-\frac{6848 \pi}{105}+\left(\frac{37664}{315}+\frac{17099 \tilde{Y}_0}{315}\right) q
\right] \ln \! \left(x \right)
+\frac{\pi (1636946+129489 \tilde{Y}_0 +2249053 \tilde{Y}_0^{2}) q^{2}}{12096} \cr
&&
+\left(-\frac{24475729}{362880}+\frac{12219385727}{34836480} \tilde{Y}_0 
-\frac{150072331}{362880} \tilde{Y}_0^{2}-\frac{32965552127}{34836480} \tilde{Y}_0^{3}\right) q^{3} \cr
&&
+\left(-\frac{33}{1280}+\frac{146033}{143360} \tilde{Y}_0 -\frac{801}{640} \tilde{Y}_0^{2}-\frac{73}{160} \tilde{Y}_0^{3}-\frac{537}{256} \tilde{Y}_0^{4}+\frac{1061}{4096} \tilde{Y}_0^{5}\right) q^{5}. \cr
\end{eqnarray*}

\section{Convergence of PN formulas for $E$ and $L$}
\label{sec:convergence-EL}

In a similar manner to Eq.~\eqref{eq:PN_diff_dotEL}, we check the convergence
of the PN formulas of the specific energy and angular
momentum, Eqs.~\eqref{eq:En} and~\eqref{eq:Ln}, by introducing the difference
\begin{equation}
\Delta E^{(n)} \equiv \left| E^{(n)} - E^{(n-1)} \right|, \quad
\Delta L^{(n)} \equiv \left| L^{(n)} - L^{(n-1)} \right|.
\label{eq:PN_diff_EL}
\end{equation}

\begin{figure}[!ht]
\includegraphics[bb=0 0 853 518, width=0.8\linewidth]{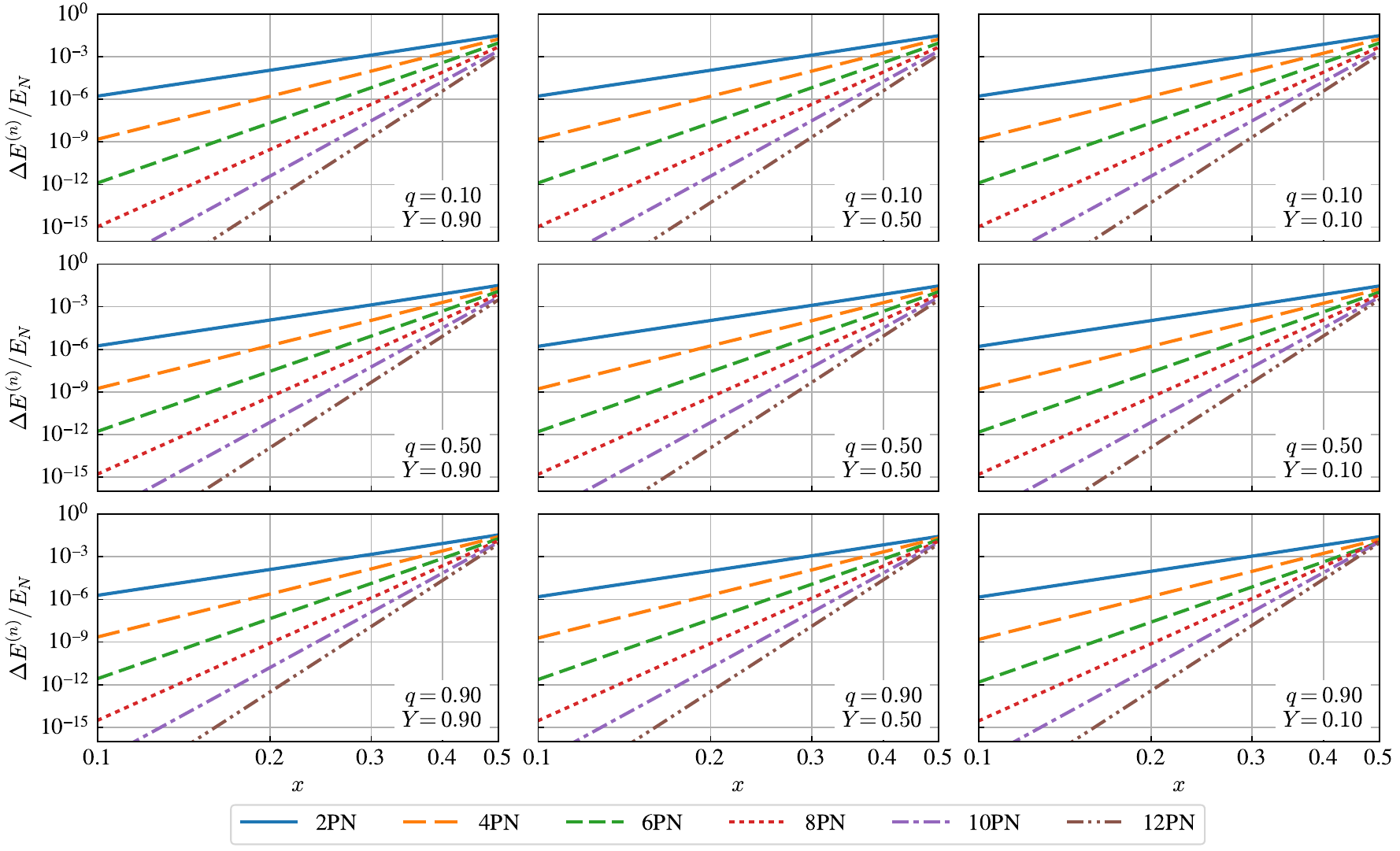}
\caption{Contribution of each correction term in the PN expansion of the specific energy.
We plot $\Delta E^{(n)}$ normalized by the Newtonian energy, $E_N$, as functions of $x$
for several sets of $(q, Y)$. 
In the same reason as 
Fig.~\ref{fig:convergence-dotE}, we show only the 2PN, 4PN, 6PN, 8PN, 10PN and 12PN 
corrections in this figure.}
\label{fig:convergence-E}
\end{figure}

\begin{figure}[!ht]
\includegraphics[bb=0 0 853 518, width=0.8\linewidth]{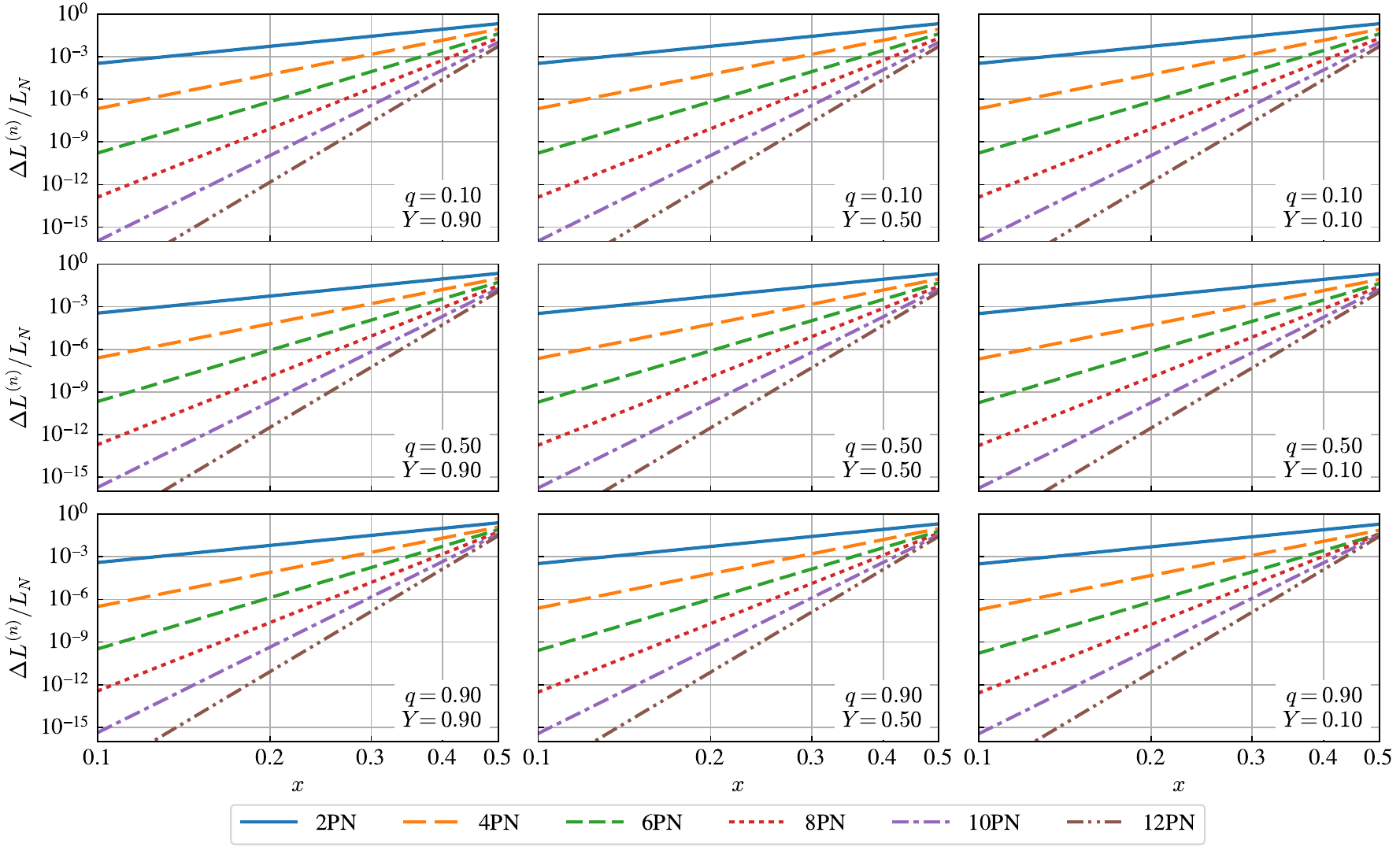}
\caption{
Contribution of each correction term in the PN expansion of the specific angular momentum.
We plot $\Delta L^{(n)}$ normalized by the Newtonian term, $L_N$, as functions of $x$
for several sets of $(q, Y)$. 
In the same reason as 
Fig.~\ref{fig:convergence-dotE}, we show only the 2PN, 4PN, 6PN, 8PN, 10PN and 12PN 
corrections in this figure.}
\label{fig:convergence-L}
\end{figure}

In Figs.~\ref{fig:convergence-E} and~\ref{fig:convergence-L}, we show $\Delta E^{(n)}$
and $\Delta L^{(n)}$ as functions of $x$ for several sets of $(q, Y)$. In these plots,
$\Delta E^{(n)}$ and $\Delta L^{(n)}$ are normalized by the leading (Newtonian) order terms,
\begin{equation*}
E_N \equiv 1 - \frac{1}{2}x^2, \quad
L_N \equiv \frac{Y}{x}.
\end{equation*}
For all cases, the difference decreases uniformly in $x < x_\textrm{fin}$
when the order of $x$ increases although the decrease is very slow near
$x \sim x_\textrm{fin}$ for System2.
This means that the PN formulas converge well in this region.

We find a weak dependence on $q$ in Figs.~\ref{fig:convergence-E} and~\ref{fig:convergence-L}:
the PN convergence gets slightly worse when $q$ increases.
On the other hand, we find no dependence on $Y$ in these plots at least with one's eyes.

\bibliography{GWphase_spherical}

\end{document}